\documentclass[a4paper,11pt]{article}
\pdfoutput=1

\usepackage{mathrsfs,graphicx,rotating,amsmath,amsfonts,mathtools,booktabs,amssymb}
\usepackage{hyperref}\usepackage{slashed}
\usepackage[nosort]{cite}
\usepackage[table,xcdraw]{xcolor}
\hypersetup{colorlinks,bookmarksopen,bookmarksnumbered,
linkcolor=blue,pdfstartview=FitH,urlcolor=rossos,citecolor=verde}
\allowdisplaybreaks

\newcommand{\mio}[1]{}

 \newcommand{\doi}[1]{\href{http://dx.doi.org/#1}{[link]}}

\allowdisplaybreaks
\usepackage{multicol}
\usepackage{color}
\definecolor{rosso}{cmyk}{0,1,1,0.4}
\definecolor{rossos}{cmyk}{0,1,1,0.55}
\definecolor{rossoc}{cmyk}{0,1,1,0.2}
\definecolor{blu}{cmyk}{1,1,0,0.3}
\definecolor{blus}{cmyk}{1,1,0,0.6}
\definecolor{bluc}{cmyk}{1,1,0,0.1}
\definecolor{verde}{cmyk}{0.92,0,0.59,0.25}
\definecolor{verdec}{cmyk}{0.92,0,0.59,0.15}
\definecolor{verdes}{cmyk}{0.92,0,0.59,0.4}

\newcommand{\GeV}{\,{\rm GeV}}

\def\circa#1{\,\raise.3ex\hbox{$#1$\kern-.75em\lower1ex\hbox{$\sim$}}\,}

\newcommand{\beq}{\begin{equation}}
\newcommand{\eeq}{\end{equation}}

\newcommand{\bea}{\begin{eqnarray}}
\newcommand{\eea}{\end{eqnarray}}
\newcommand{\be}{\begin{equation}}
\newcommand{\ee}{\end{equation}}

\newcommand{\largeN}{\frac{N^2}{16\pi^2}}

\font\tenrsfs=rsfs10 at 12pt
\font\sevenrsfs=rsfs7
\font\fiversfs=rsfs5
\newfam\rsfsfam
\textfont\rsfsfam=\tenrsfs
\scriptfont\rsfsfam=\sevenrsfs
\scriptscriptfont\rsfsfam=\fiversfs

\def\circa#1{\,\raise.3ex\hbox{$#1$\kern-.75em\lower1ex\hbox{$\sim$}}\,}
\makeatletter

\def\hhref#1{\href{http://arxiv.org/abs/#1}{arXiv:#1}} 
 
\def\art{\@ifnextchar[{\eart}{\oart}}
\def\eart[#1]#2#3#4#5#6{{\rm #2}, {\em #3 \bf #4} {\rm (#6) #5} ({\em #1})}

\def\article{\@ifnextchar[{\earticle}{\oarticle}}
\def\oarticle#1#2#3#4#5#6{{\rm #1}, {\em ``#6''}, {\rm #2 #3 (#5) #4}}
\def\earticle[#1]#2#3#4#5#6#7{{\rm #2}, {\em ``#7''}, {\rm #3 #4 (#6) #5}  [\hhref{#1}]}
\def\hepart[#1]#2{{\rm #2, \em#1}}
\def\heparticle[#1]#2#3{#2, {\em ``#3''} [\hhref{#1}]}

\newcounter{alphaequation}[equation]
\def\thealphaequation{\theequation\hbox to
0.6em{\hfil\alph{alphaequation}\hfil}}
\def\eqnsystem#1{
\def\@eqnnum{{\rm (\thealphaequation)}}
\def\@@eqncr{\let\@tempa\relax \ifcase\@eqcnt \def\@tempa{& & &} \or
  \def\@tempa{& &}\or \def\@tempa{&}\fi\@tempa
  \if@eqnsw\@eqnnum\refstepcounter{alphaequation}\fi
\global\@eqnswtrue\global\@eqcnt=0\cr}
\refstepcounter{equation} \let\@currentlabel\theequation \def\@tempb{#1}
\ifx\@tempb\empty\else\label{#1}\fi
\refstepcounter{alphaequation}
\let\@currentlabel\thealphaequation
\global\@eqnswtrue\global\@eqcnt=0 \tabskip\@centering\let\\=\@eqncr
$$\halign to \displaywidth\bgroup \@eqnsel\hskip\@centering
$\displaystyle\tabskip\z@{##}$&\global\@eqcnt\@ne
\hskip2\arraycolsep\hfil${##}$\hfil& \global\@eqcnt\tw@\hskip2\arraycolsep
$\displaystyle\tabskip\z@{##}$\hfil
\tabskip\@centering&\llap{##}\tabskip\z@\cr}
\def\endeqnsystem{\@@eqncr\egroup$$\global\@ignoretrue} \makeatother

\begin{document}

\vspace{1cm}

\begin{center}
{\huge\bf 
Gravitational Waves from Supercool Axions}\\

\bigskip
\bigskip
\bigskip

{\bf Luigi Delle Rose, Giuliano Panico, Michele Redi, Andrea Tesi}  
\\[7mm]

{\it  INFN, Sezione di Firenze and Department of Physics and Astronomy, University of Florence,\\
Via G. Sansone 1, 50019 Sesto Fiorentino, Italy}\\[1mm]

\begin{figure}[t]
\begin{center}
\label{fig:gluinos}
\end{center}
\end{figure}

\vspace{2cm}
{\large\bf Abstract}
\begin{quote}\large
We study the dynamics of the Peccei--Quinn (PQ) phase transition for the QCD axion.
In weakly coupled models the transition is typically second order except in the region of parameters where the PQ symmetry is broken through the Coleman--Weinberg mechanism.
In strongly coupled realizations the transition is often first order.  We show examples where the phase transition leads to strong supercooling lowering 
the nucleation temperature  and enhancing the stochastic gravitational wave signals. The models predict a frequency peak in the range 100-1000 Hz with an amplitude that is already within the sensitivity of LIGO and can be thoroughly tested with future gravitational wave interferometers. 
\end{quote}

\bigskip
\bigskip
\bigskip

\thispagestyle{empty}
\end{center}

\vfill
\noindent\line(1,0){188}
{\scriptsize{ \\ E-mail:\texttt{ \href{mailto:luigi.dellerose@fi.infn.it}{luigi.dellerose@fi.infn.it}, \href{giuliano.panico@unifi.it}{giuliano.panico@unifi.it}, \href{mailto:michele.redi@fi.infn.it}{michele.redi@fi.infn.it}, \href{andrea.tesi@fi.infn.it}{andrea.tesi@fi.infn.it}}}}
\newpage
\newpage

\tableofcontents

\section{Introduction}

Cosmological phase transitions are of particular interest when they are first order as they correspond to 
dramatic changes of the degrees of freedom of the theory and lead, among other things, to the production of gravitational waves (see Ref.~\cite{Caprini:2015zlo} for a review). 
The Standard Model (SM) predicts the existence of two phase transitions, the QCD and the electroweak one, which however are not first order. 
Therefore the detection of a signal compatible with a first order phase transition would be a sharp evidence of physics beyond the SM.

Perhaps, the most strongly motivated phase transition beyond the SM is the Peccei--Quinn (PQ)~\cite{Peccei:1977hh} phase transition associated to the QCD axion. 
The axion solution of the strong CP problem~\cite{Peccei:1977hh,Weinberg:1977ma,Wilczek:1977pj} requires the existence of a U(1)$_{\rm PQ}$ global symmetry anomalous under QCD  that is spontaneously broken at
a scale $f_a> 10^9$ GeV. Even more compelling is the situation in which the axion provides the whole Dark Matter abundance.
This possibility is connected to how the PQ symmetry is restored in the early Universe. Depending on the scale of inflation
and reheating, two  rather different scenarios emerge, in which the initial value of the axion in the visible Universe is either constant or scans all possible values. 

In this work we study in detail the PQ phase transition in several scenarios. In the past this question raised limited interest,
mainly because the low-energy axion phenomenology relevant for experiments is independent 
from the nature of the phase transition. Today, however, the situation is rather different,
since the possibility to test gravitational-wave (GW) signals opened up a powerful way to test the PQ dynamics.

In order to get a detectable signal, we need to assume that the PQ transition took place after inflation. If this is not the
case any possible signal of the unbroken PQ phase is completely erased. Moreover we will mostly focus on scenarios in which
the axion can provide the whole dark matter. This happens for $f_a \sim 10^{11}\,{\rm GeV}$, which we use as our
main benchmark \cite{Preskill:1982cy,Abbott:1982af,Dine:1982ah}.

We present concrete examples where the PQ phase transition is first order and a detectable GW-signal is produced. The way this works is as follows. A first order phase transition is automatically obtained when the theory is approximately conformal. The PQ symmetry is  broken dynamically through the Coleman--Weinberg mechanism. The small deviation from conformality implies  a suppression of the transition probability, so that a large amount of supercooling is generic and calculable~\cite{Witten:1980ez}. 
Supercooling implies that bubble collisions take place in the vacuum and increases the duration of the phase transition, thus enhancing the GW signal.

In this paper we will show that the scenario above can be realized  both at weak and at strong coupling.
In the weakly coupled case approximate scale invariance is achieved at the price of tuning, since we have to require small or vanishing mass terms for the scalar fields. This can be considered unnatural if there is heavy new physics  coupled to the scalar sector.
In the strongly coupled scenario naturalness depends on the unknown dimension of the deformations of the CFT. Moreover higher dimensional operators can spoil the `quality' of the PQ solution (see \cite{Kamionkowski:1992mf}) and modify the dynamics of the phase transition. However we notice that, for the phase transition itself, operators suppressed by the Planck scale will not affect our conclusions.

While our main focus are scenarios where a detectable GW signal is produced, we take the opportunity to study the nature 
of the PQ phase transition in other popular models. For instance, we show that in the minimal KSVZ model~\cite{KSVZ1,KSVZ2} the phase transition is always second order,
while in composite axion models based on renormalizable gauge theories~\cite{Choi:1985cb} the phase transition is first order 
but nucleation proceeds rapidly with small supercooling, leading to a suppressed GW signal.

The paper is organized as follows. In section 2 we consider weakly coupled theories where the axion is an elementary scalar.
After showing that in KSVZ models the phase transition is second order, we show that in theories with massless scalars the PQ symmetry is broken 
through the Coleman--Weinberg mechanism leading to a first order phase transition. In section 3 we discuss the realization of this mechanism in 
spontaneously broken strongly-coupled conformal field theories and their dual Randall--Sundrum-like incarnations. 
In both cases we find that the GW signal is within the reach of present and future Earth-based GW interferometers such as LIGO/VIRGO and the Einstein Telescope (ET).

\section{Elementary axions}

In this section we study axion models that are described in terms of fundamental scalars at all energy scales. In this context the axion is the phase of an elementary complex scalar field whose VEV is $f_a$. We analyze two classes of theories: models of KSVZ type, and models in which the breaking of the PQ symmetry is radiatively induced.\footnote{We do not consider DFSZ models because they have domain-wall number greater than one so that in the minimal scenario PQ symmetry should be broken during inflation.}

\subsection{KSVZ-type}\label{sec:KSVZ}

In the simplest realization of the QCD axion one introduces a complex scalar field $X$ and colored fermions charged under a new U(1) PQ symmetry. 
Taking into account the Higgs doublet, the most general renormalizable lagrangian includes a potential
\begin{equation}\label{PQ-potential}
V= -\mu^2 |H|^2 +\lambda |H|^4+\lambda_{X H} |X|^2 |H|^2 + \lambda_X (|X|^2-f^2/2)^2\,,\quad X=\frac{\phi}{\sqrt{2}}\exp(ia/f_a)\,.
\end{equation}

In order to discuss whether this model displays a first order phase transition it is convenient to start from the case where the portal coupling is vanishing, $\lambda_{XH}=0$. In this situation the Higgs and the PQ field will follow separate dynamics, and it is easy to see that PQ transition is of second order. The reasons for this are several: $i)$ radiative corrections induced by the self-coupling $\lambda_{X}$ at zero temperature do not generate other minima in the regime where a perturbative expansion applies; $ii)$ temperature corrections are not able to modify the potential from the `mexican hat' shape, since for $X\approx 0$ there are no light bosonic states that could induce a temperature-dependent barrier between the origin and the true minimum; $iii)$ the potential for $X$ is always well approximated by $m_X(T)^2 |X|^2 +\lambda(T) |X|^4$, and no maxima away from the origin are expected. This conclusion agrees with the analysis of \cite{smash}.

Then one might wonder if  departure from second order can be achieved by exploiting the inevitable coupling of the field $X$ to the Higgs at the renormalizable level, especially in the regime where $\lambda_{XH}\gg \lambda_X$. We argue in the following that this is not the case.

When the Higgs sector is brought into play, we need to ensure that the electroweak VEV and the Higgs mass are reproduced. 
Therefore, in addition to the stability of the potential, which is guaranteed if $\lambda,\lambda_X>0$ and $\lambda_{X H}>-2\sqrt{\lambda_X\lambda}$,
we have two additional constraints: $\mu^2$ has to be tuned against the contribution from the portal, and a correlation between the quartics is needed to ensure that $\lambda_h=2M_h^2/v^2$. In the minimum $(v,f_a)$, where both the electro-weak and PQ symmetries are broken, by integrating out the massive singlet $\phi$ we get at leading order in $v^2/f^2$
\be\label{matching-KSVZ}
\frac{M_h^2}{2}=\mu^2 -\frac{\lambda_{XH}f_a^2}{2},\qquad \quad \lambda_h = \lambda -\frac{\lambda_{XH}^2}{4\lambda_X}\,.
\ee
These matching conditions strongly constrain the possible sizes of our parameters. In addition, this configuration is the deepest minimum if $\mu^2>\lambda_{XH}f^2/2$, $\lambda_{X H}^2<4\lambda \lambda_X$ and $\mu^2<2\lambda\lambda_X f^2/\lambda_{X H}^2$ (for a positive portal coupling). Notice that if the first two conditions are satisfied, the third is implied. 

Given the separation of scales $v\ll f$, it is a good approximation to study the effective potential along the direction $h\approx 0$.\footnote{This is a good approximation because the Higgs receives
larger thermal masses than $\phi$. Even neglecting the gauge and Yukawa contributions, one can see that the singlet becomes unstable at temperatures higher than the Higgs field when $\lambda_{X H}$ is not too large. 
} The needed departure from a pure `mexican hat' potential for $X$ can arise from radiative and thermal contributions. Both the CW one-loop effective potential and the thermal corrections depend on the masses of the fields involved in the dynamics. Building on the previous discussion it is necessary to invoke a hierarchy $\lambda_{XH}\gg \lambda_X$, and in this limit the dominant effects comes from the four degrees of freedom of the Higgs doublet with effective mass $\lambda_{X H}(\phi^2-f^2)/2$. Neglecting terms of order $\lambda_X^2$, the potential depends only on one variable
\be\label{ksvz-limit}
V_{\rm CW+tree+thermal}\approx \frac{1}{2}\frac{\lambda_{X H} T^2}{6} s^2+  \frac{\lambda_X}{4} s^4 + \frac{\lambda_{X H}^2 s^4}{64\pi^2}\log\bigg(\frac{\lambda_{X H}}{2 \bar \mu^2} \big|s^2\big|\bigg) \,,\quad s^2\equiv \phi^2-f^2\,,
\ee
where we take the real part of the logarithm and $\bar\mu=\mu e^{3/4}$ and we included the leading high-temperature corrections (see section \ref{sec:weakcoupling}). 
Notice that the variable $s^2$ is bounded by $ s^2\geq -f^2$. At high-temperature the minimum is at $s_{\rm min}^2\to -f^2$, while for lower temperatures it goes $ s_{\rm min}^2\to 0$.
When $\lambda_{X H}^2\gtrsim 16\pi^2 \lambda_X$ the potential deviates from a pure `mexican hat' shape and in this regime two minima coexist. However this hierarchy of couplings has to be faced with the requirement of the Higgs properties reported in eq.~\eqref{matching-KSVZ}. This implies $\lambda\gg 16\pi^2$, which is not viable phenomenologically. If the parameters are not constrained by Higgs phenomenology a larger parameter space opens up, see for example the regime discussed in Ref.~\cite{1905.00891}.

\medskip

The KSVZ example suggests how to modify the theory to find a strong first order phase transition. 
First the Higgs field should be replaced by a field that does not play a role in electroweak symmetry breaking. 
Moreover, to obtain a sizable GW signal, the system should undergo a phase of supercooling. 
This is achieved if the theory is approximately conformal since
\begin{equation}
{\rm CFT} \quad \Longrightarrow \quad \frac {S_3} T= {\rm constant}\,,
\end{equation}
where $S_3$ is the euclidean tunneling bounce action at finite temperature, which determines the false vacuum decay rate $\Gamma \sim e^{-S_3/T}$.

At weak coupling this can be realized through massless scalar theories where PQ symmetry breaking is induced by quantum corrections~\cite{Coleman:1973jx,Gildener:1976ih}.

\subsection{Radiative PQ breaking}\label{sec:weakcoupling}

To realize this scenario we consider theories of (approximately) massless scalars some of which are charged under the PQ symmetry.
The tree level potential is given by
\begin{equation}\label{model}
V=\frac{\lambda_{ijkl}}{4} \phi_i \phi_j\phi_k \phi_l.
\end{equation}
It is well known that such theories undergo spontaneous symmetry breaking. The way this works is as follows.
Renormalization group equations imply generically that a linear combination of couplings vanishes at some scale $\Lambda$~\cite{Gildener:1976ih}. Starting from this scale one can identify a ``flat'' direction in the scalar vacuum manifold parametrized by a unit vector $\vec{n}$ and spanned by a field $\sigma$ ($\vec{\phi}=\vec{n} \sigma$).
Using perturbation theory with a renormalization scale $\mu = \Lambda$, the tree-level effective potential
along the $\sigma$ direction identically vanishes.
The whole dynamics of $\sigma$ is therefore controlled by radiative effects, and can  be described in terms
of an effective quartic coupling
\be
\lambda_{\rm eff}(\mu) = \lambda_{ijkl}(\mu)n_i n_j n_k n_l\,, \quad \mathrm{with}\quad \lambda_{\rm eff}(\Lambda)=0\,.
\ee

At 1-loop the Coleman-Weinberg effective potential is given by
\begin{equation}\label{CW}
V_{\rm  eff}(\sigma)\approx  \frac{\beta_{\lambda_{\rm eff}}}4 \sigma^4 \left(\log \frac {\sigma}{f}-\frac  {1} {4}\right)\,,
\end{equation}
where $\beta_{\lambda_{\rm eff}}$ is the $\beta$ function associated to the effective quartic coupling. We see that if $\beta_{\lambda_{\rm eff}}>0$ the potential has a minimum at $f$, which is close to the scale where the effective quartic becomes negative. 
At the minimum, $\sigma$ has a mass $m_\sigma^2=\beta_{\lambda_{eff}}f^2$.
This is the usual radiative symmetry breaking \`a la Coleman-Weinberg. 
\medskip

\begin{figure}[t]
\centering
\includegraphics[width=1\textwidth]{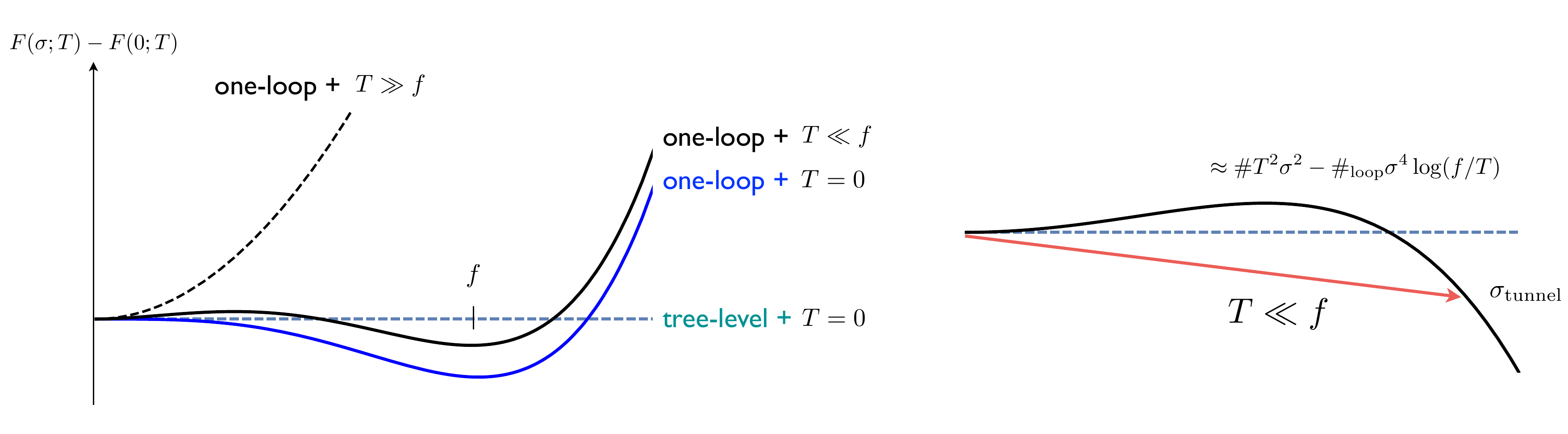} 
\caption{\it{On the left, free energy along $\sigma$ for different values of the temperature. 
For $T\gg f$ the origin is the only minimum so that the symmetry is unbroken.
At lower temperature a new minimum develops separated by a barrier of size proportional to $T$.
The Universe is trapped in the false vacuum up to low temperatures where the relevant
region for tunneling is well described by temperature dependent quartic potential (right panel).}}
\label{fig:cartoon}
\end{figure}

As noted in Ref.~\cite{Witten:1980ez}, due to the extremely shallow potential, finite temperature effects have a dramatic impact on the free-energy along the flat direction,
see also \cite{Iso:2017uuu,Hambye:2018qjv} for recent work.
In particular, for any $T>0$ thermal corrections induce a positive curvature at the origin, making $\sigma=0$ a metastable vacuum.

The dynamics of the system at finite temperature is described by the free-energy,
which, in a theory of weakly-coupled scalars, is  given by\footnote{We recall that the thermal functions are given by
\be
J_{B/F}(y^2)=\int_0^\infty dt\, t^2 \log(1 \mp \exp(-\sqrt{t^2+y^2}))\,.
\ee
Their expansions at high temperature $|y^2|\ll 1$ are
\begin{eqnarray}
J_{B}(y^2)&=&-\frac{\pi^4}{45}+\frac{\pi^2}{12}y^2 -\frac{\pi}{6}y^3 -\frac{y^4}{32}\log(\frac{y^2}{a_B}), \quad  a_B=\pi^2 e^{3/2-2\gamma_E},\\
J_F(y^2)&=&  \frac{7\pi^4}{360}-\frac{\pi^2}{24}y^2-\frac{y^4}{32}\log(\frac{y^2}{a_F}),\quad a_F=16\pi^2 e^{3/2-2\gamma_E}\,.
\end{eqnarray}}
\be
F(\sigma; T) = \frac{T^4}{2\pi^2}\sum_i^{N} J_B \left(\frac{m_i^2(\sigma)}{T^2}\right) + V_{\rm eff}(\sigma) + V_0\,,
\ee
where $V_0=- V_{\rm eff}( \sigma =f)$ is included to eliminate the cosmological constant on the global minimum.

Due to the absence of tree-level mass terms, close to the origin the thermal function can be expanded in the high-temperature limit, since all degrees of freedom are massless there. Notice that for a large number ($N$) of scalar fields coupled to $\sigma$, the variable $\sigma$ has an overlap of order $1/\sqrt{N}$ with each of them. It then follows that the $\sigma$-dependent masses of the light degrees of freedom are of order $m_i \sim \hat g \sigma/\sqrt{N}$, where $g \sim \sqrt{\lambda}$ is the typical interaction strength in eq.~\eqref{model}. Therefore for $T\gtrsim \hat g\sigma/\sqrt{N}$ a formally high-temperature expansion is reliable.

The shallowness of the potential implies that the critical temperature $T_c$ at which $F$ develops two minima is parametrically smaller than $f$, namely
\be
\frac{T_c^4}{f^4}\approx \frac 1 {8\pi^2}\frac {45} {(N_H-N_L)} \beta_{\lambda_{\rm eff}}\,,
\ee
where $N_H$ and $N_L$ are the number of light degrees of freedom at the origin and at the global minimum.

Working in the limit of temperatures smaller than $f$ but larger than $\hat g \sigma/\sqrt{N}$, the free-energy can be approximated as
\be\label{pot-witten}
F(\sigma; T) \approx - N \frac{\pi^2}{90} T^4+ a \frac{ \hat{g}^2 T^2}{24}\sigma^2  - \frac{\beta_{\lambda_{\rm eff}}}{4}\sigma^4 \log\left(\frac{M}{T}\right)+V_0\,,
\ee
where $M\sim \hat{g} f/\sqrt{N}$ is the typical mass of the heavy scalars around the minimum. Here $a$ is an O(1) coefficient at large $N$.
Notice that the approximated potential is just a quartic polynomial and no logarithmic dependence on the field $\sigma$
is present. This feature is a consequence of a cancellation between the logarithmic term in the CW potential and the
logarithmic piece in the expansion of the thermal functions.

The potential in eq.~(\ref{pot-witten}) has a minimum at the origin and a barrier whose size is roughly given by $\sigma_{\rm barrier} \approx \sqrt{a/6}(\hat g/\sqrt{\beta_{\lambda_{\rm eff}} \log(M/T)}) T$. Thanks to the logarithm, which becomes sizable for large supercooling, the barrier extends over a region where the approximation of high-temperature is still reliable~\cite{Witten:1980ez}.
At sufficiently low temperature, the field $\sigma$ will then tunnel towards the true minimum and acquire a vacuum expectation value. Thanks to its overlap with the original variable $\phi_i$, 
this also breaks the PQ symmetry. An exemplificative picture of the free energy of the system and of the impact of the thermal corrections to the vacuum structure is shown in figure \ref{fig:cartoon}. Also shown (right panel) is the parameterization of the barrier in the high-temperature approximation and the corresponding field tunneling.

\bigskip

Using the approximation in eq.~(\ref{pot-witten}), the tunneling rate is determined by minimizing the bounce action,
\be
\frac{S_3}{T}\approx\frac{4\pi}{T}\int_0^\infty r^2 dr \bigg[ \frac{1}{2}\left(\frac{d\sigma}{dr}\right)^2 + \frac{m^2(T)}{2}\sigma^2-\frac{\lambda_4(T)}{4}\sigma^4\bigg]\,
\ee
which is subject to the conditions $\sigma'(0)=0$ and $\sigma(\infty)=0$. Here $m^2(T)=a \hat g^2 T^2/12$  and $\lambda_4(T)=\beta_{\lambda_{\rm eff}}\log(M/T)$. 
This is just the bounce for a potential with a positive quadratic term and a negative quartic for which the exact result is $S_3\approx 18.897 m(T)/\lambda_4(T)$ \cite{parisi}.
In our case, parametrizing $16\pi^2\beta_{\lambda_{\rm eff}}=b_{\rm eff}{\hat g}^4$, the full result can be expressed as
\begin{equation}\label{witten}
\frac {S_3}T
\approx \frac {A_3}{\log(M/T)}\,,~~~~~~~~~~~~~~~A_3 =  \frac{861.43}{\hat g^3 }\frac{\sqrt{a}}{b_{\rm eff}}\,.
\end{equation}

The bounce action is large at weak coupling due to the approximate scale invariance of the theory and decreases logarithmically
at low temperatures.
This generically implies a phase of supercooling, since the tunneling rate,
\be\label{gamma}
\Gamma \simeq T^4 \bigg(\frac{S_3/T}{2\pi}\bigg)^{\frac{3}{2}} \exp(-S_3/T)\,,
\ee
is exponentially suppressed for a large range of temperatures. Therefore, when the temperature drops below the critical one, the vacuum energy of the false minimum begins to dominate over the energy density of radiation. The Universe then enters the so-called phase of supercooling, where Hubble becomes constant at a value $H_I$ and the temperature starts to drop exponentially $T\sim e^{-H_I t}$. This phase ends when the value of the bounce action decreases enough to allow nucleation of bubbles of true vacuum.

The nucleation temperature is commonly defined as when the time-integrated probability to enucleate one bubble per Hubble volume equals one (see Ref.s \cite{Ellis1,Ellis2} for more details). By exploiting the exponential growing of the tunneling rate for the relevant temperatures, one can approximate the nucleation condition by the simpler relation
\be\label{Tn-all}
\frac{\Gamma}{H^4}=q,\quad q\geq 1\quad\mathrm{(usually\ taken\ to\ be\ 1)}\,.
\ee
By using eq.~\eqref{gamma}, we obtain the following condition for the nucleation temperature $T_n$
\be\label{eq-Tn}
\frac{S_3}{T_n}- \frac{3}{2}\log\left(\frac{S_3}{2\pi T_n}\right)=4\log  \frac{T_n}{H_I} -\log q\,.
\ee
During supercooling $H_I$ is given by,
\begin{equation}
H_I^2=\frac {V_0}{3 M_{\rm Pl}^2}= \frac{\beta_{\lambda_{\rm eff}}}{48} \frac{f^4}{M_{\rm Pl}^2}\,.
\label{eq:HI}
\end{equation}
where $M_{\rm Pl}=2.4\times 10^{18} \GeV$. 

It is worth emphasizing that in the case of large supercooling it is inaccurate to simply assume $S_3/T\approx 4\log (M_{\rm Pl}/T_c)$. Since the temperature can drop significantly, the right-hand side of the nucleation condition in eq.~(\ref{eq-Tn}) can become quite small, requiring to go to even smaller values of the bounce action. Notice also that we assumed a constant Hubble value, neglecting the exponentially-decreasing radiation component.

Using (\ref{witten}) the nucleation temperature is found to be
\be\label{Tn}
T_n \approx \sqrt{M H_I} \exp\bigg(\frac{1}{2}\sqrt{-A_3+\log^2(M/H_I)}\bigg)\,.
\ee
The argument of the square root becomes negative for sufficiently small $\hat g$, implying that the nucleation temperature has a limiting value
\begin{equation}
T_n^{\rm min}= \sqrt{M H_I}\sim 0.1 f \bigg(\frac {f}{M_{\rm Pl}}\bigg)^{\frac{1}{2}} \,.
\end{equation}
For temperatures above this value we however expect a sizable tunneling rate and the completion of the phase transition via thermal tunneling.

Within the above approximation, one can also compute the logarithmic derivative of the tunneling rate, which is one of the relevant parameters to determine the GW spectrum.
Its expression is given by
\be\label{beta-H}
\frac{\beta}{H}=-4 + T \frac{\partial (S_3/T)}{\partial T}\bigg|_{T_n}= -4 + \frac{1}{\log(M/ T_n)} \frac {S_3}{T}\bigg|_{T_n}\,,
\ee
where $T_n$ and $\hat S_3(\hat g)$ are related by eq.~\eqref{Tn}. Since $T_n\ll T_c< \lambda f$  we see immediately that $\beta$ can become $O(1)$, in which case the power spectrum of the GWs is maximized.

\bigskip
Apart from tunneling at finite temperature, nucleation of true vacuum bubbles can also be driven by 4d bounces. If the ${\rm O}(4)$ bounce action $S_4$
is smaller than $S_3/T$, quantum effects can lead to a faster nucleation rate.
Repeating the same steps as above we find\footnote{Strictly speaking the O(4) bounce does not exist 
for $M>0$ because no trajectory starting at finite $\phi$ reaches exactly the false vacuum $\phi=0$. 
A small modification of the potential such as a logarithmic modulation produces a solution whose action is 
roughly equal to the theory with $M=0$.}
\be
S_4 =\frac{2\pi^2}{\lambda_4}\int_0^\infty z^3 dz \bigg[ \frac{1}{2}\big(\frac{d\hat\sigma}{dz}\big)^2 + \frac{1}{2}\hat\sigma^2-\frac{1}{4}\hat\sigma^4\bigg] \approx \frac{25}{\lambda_4} \equiv \frac{\hat S_4 (\hat g)}{\log(M/T)}\,,
\ee
This action is larger than $S_3/T$ as long as $M(T)/T<1$. The expression for the tunneling rate is
\be\label{gamma4}
\Gamma \simeq \frac 1 {R^4}\bigg(\frac{S_4}{2\pi}\bigg)^2 e^{-S_4}
\ee
where $R$ is the size of the bubble. Since $R \lesssim 1/T$, and $S_4 < S_3/T$, we find that the thermal nucleation rate
always dominates.

\bigskip

The above results show that, for small enough coupling, the thermal transition never completes.
Does this suggest that the inflationary epoch lasts up to arbitrarily small temperatures? 
The answer is no, fortunately, because when the temperature of the thermal bath drops below the Hubble scale in the false vacuum  the computation of the tunneling rate should be modified to take into
account the de Sitter curvature~\cite{1503.05193,1706.00792}. This happens after a number of e-foldings
\begin{equation}
N_{\rm max} \approx \log  \frac{T_c}{H_I}\approx \log \bigg[\frac {M_{\rm Pl}}{f} \left(\frac {90}{N_H \pi^2 \beta_{\lambda_{\rm eff}}}\right)^{1/4}\bigg]\,.
\end{equation}
Since $N_{\rm max}\approx 15$, the model is consistent with the CMB power spectrum.

At sufficiently low temperatures, the height of the barrier will be smaller than $H_I$. In this regime quantum de Sitter fluctuations in the false vacuum, whose variance is $\delta \sigma =H_I/(2\pi)$, will allow the field to reach its true minimum. We do not study this regime further  in this work.

\bigskip

After the completion of the phase transition the Universe is reheated at a temperature
\begin{equation}\label{temperature-inflation}
T_{\rm RH}= T_I\, {\rm min}\left(1, \frac{\Gamma}{H_I}\right)^{1/2}\,,\qquad\quad T_I=\left(\frac {\beta_{\lambda_{\rm eff}}} {16\pi^2}\frac {30}{N_L} \right)^{1/4} f\,.
\end{equation}
The reheating is controlled by the coupling to PQ fermions. 
In a large range of parameters the decay rate is fast, so that $T_{\rm RH}\approx T_I$. It is however possible to suppress the decay rate by considering small Yukawa couplings. In this case, the PQ sector gets reheated first and afterwards the energy is transferred to the SM.

\subsubsection{An explicit realization}\label{sec:explicit_realization}

We now discuss an explicit implementation of radiative PQ breaking. We consider a  pair of complex scalar fields, $S$ and $X$, neutral under the SM and coupled to colored vectorial fermions $Q$ and $Q_c$. As an example, we assume $S$ to be neutral under PQ and we do not include tree level couplings to the Higgs doublet. The corresponding Lagrangian is given by
\be
 \mathscr{L}= -\frac{F^2}{4g^2}+ |D_\mu S|^2+|\partial_\mu X|^2 - V+  (y X Q Q^c + h.c)\,, \quad V= \lambda_S |S|^4 + \lambda_X |X|^4 + \lambda_{XS} |S|^2 |X|^2\,.
\ee
The fermions $Q,Q^c$ are assumed to transform in the fundamental and anti-fundamental representations of color, and to have hyper-charge equal to $-2/3$ or $1/3$. In this way  the domain wall number is equal to one, and they can decay by mixing with the right-handed quarks. We also included a possible ${\rm U}(1)_S$ gauge symmetry,
with a small coupling strength $g$, under which only the $S$ field is charged. Very similar type of models have been considered in the context of the electro-weak phase transition
\cite{Hambye:2013sna,Iso:2017uuu,Hambye:2018qjv}. 

The tree-level scalar potential has a flat direction for $\lambda_{X S}=-2\sqrt{\lambda_S \lambda_X}$ parametrized by
\begin{equation}
(S\,,X)=(\sin \alpha\,, \cos \alpha) \frac {\sigma}{\sqrt{2}}\,,\qquad\quad \sin^2 \alpha=\frac{\sqrt{\lambda_{X}}}{\sqrt{\lambda_X}+\sqrt{\lambda_S}}\,.
\end{equation}
Along this trajectory the fields have masses
\begin{equation}
M_\tau= (4\lambda_X \lambda_S)^{1/4} \sigma\,,\qquad\quad M_A= g \sin \alpha\, \sigma\,,\qquad\quad M_Q=y \cos  \alpha\, \frac{\sigma}{\sqrt{2}}\,,
 \end{equation}
where $\tau$ is the radial direction orthogonal to $\sigma$.

Assuming $\lambda_{X S}+2\sqrt{\lambda_S \lambda_X}$ to vanish at a scale $\Lambda$, the 
Coleman-Weinberg effective potential along the direction $\sigma$ reads
 \be
V_{T=0}(\sigma) =\frac {2 \lambda_S\lambda_X+\frac 3 2 g ^4 \sin^4\alpha-\frac 3 2 y ^4 \cos^4\alpha} {16\pi^2}\, \sigma^4 \left(\log \frac{\sigma}{f} - \frac{1}{4}\right)\,,
\label{eq:CWpot}
\ee
 where, for $y$ sufficiently small, we traded the scale $\Lambda$ for the minimum of the potential at $f$. At the minimum, $\sigma$ has a loop suppressed mass, while the phase of $X$ is exactly massless being an exact Nambu-Goldstone boson up to QCD anomalies, the axion.
Note that the axion decay constant is $f_a=  f \cos\alpha$ .

 \paragraph{Thermal corrections and 3D bounce:}
 
Adding finite temperature corrections, the free-energy along the flat direction becomes\footnote{
Around the origin all fields are massless so  the whole potential is dominated by the thermal effects. However, not for all values of the quartic couplings the origin is a minimum at high temperatures. For example when $g=0$ we find instabilities for $\lambda_S>4\lambda_X$. On the contrary we find that in presence of the gauging $\lambda_S/\lambda_X$ can be arbitrarily small, allowing to align the flat direction with the $S$ axis.
}
\be
V(\sigma)= \frac{T^4}{2\pi^2} \left[J_B \bigg(\frac{M_\tau^2}{T^2}\bigg)+3J_B \bigg(\frac{M_A^2}{T^2}\bigg)-12J_F \bigg(\frac{M_Q^2}{ T^2}\bigg)\right]  + V_{T=0}(\sigma,\tau=0)\,.
\label{eq:finiteTpot}
\ee
In terms of the parametrization of the previous section this corresponds to
\begin{equation}
a\hat{g}^2=2 \sqrt{\lambda_S \lambda_X}+3 g^2 \sin^2 \alpha-3 y^2 \cos^2 \alpha\,,\qquad b_{\rm eff}\hat{g}^4=8\lambda_S \lambda_X+6 g^4 \sin^4 \alpha-6 y^4 \cos^4 \alpha\,.
\end{equation}
Using the previous results, the thermal bounce action can be approximated by
\be\label{eq:an_approx_1}
\frac{S_3}{T} \approx {\rm Max} \left[\frac{150}{(\lambda_X\lambda_S)^{3/4}}\,,\frac {250}{g^3\sin^3\alpha}  \right]\frac{1}{\log (M/T)}\,.
\ee
The approximate formula used to draw the gray dashed lines in all the plots of this section is the above eq.~\eqref{eq:an_approx_1}, multiplied by 0.8 that takes into account the cubic terms of the thermal potential, and where $M$ is the maximum between the mass of the radial mode or the guage boson.

In the fixed-order effective potential the perturbative expansion may break down when the scalar field explores a region in field space far from the renormalization point. 
This happens generically since the quartic couplings are defined at the scale $\Lambda$ but the bounce probes the scalar potential near the metastable minimum.
The remedy is to use the renormalization group improved CW effective potential.
Being the dynamics uniquely determined in the $\sigma$ direction, this simply amounts to the replacement, in the $\overline{\textrm{MS}}$ effective potential,
of the quartic couplings $\lambda_i$ with the corresponding running ones $\lambda_i(\mu)$ setting $\mu = \sigma$. We also assume that the flat direction is not strongly affected by the one-loop radiative corrections since in the orthogonal direction the tree-level potential dominates. 
This is exactly true in the symmetric configuration $\lambda_S = \lambda_X$ for negligible values of the gauge and Yukawa couplings.

Explicitly the 1-loop $\beta$ functions are given by
\be\label{eq:beta-functions}
\begin{array}{l}
16\pi^2 \beta_{\lambda_X} = \lambda_{XS}^2 + 20\lambda_X^2 +6y^2( 2\lambda_X - y^2)\,,\quad
16\pi^2 \beta_{\lambda_S} = \lambda_{XS}^2 + 20\lambda_S^2+ 6 g^4-12 g^2\lambda_S\\
\rule{0pt}{1.5em}16\pi^2 \beta_{\lambda_{XS}}= 4\lambda_{XS}^2+8\lambda_{XS} (\lambda_X+\lambda_S) \,,\quad
16\pi^2\beta_g={g^3}/3\,,\quad 16\pi^2 \beta_y = 5y^3/2\,.
\end{array}
\ee
From the above equations one can extract the running of $\lambda_{\rm eff}$. Assuming that it vanishes at tree level ($\lambda_{XS}=-2\sqrt{\lambda_X \lambda_S}$) one finds
\be
16\pi^2 \beta_{\lambda_{\rm eff}}=8\lambda_X\lambda_S -3 \sqrt{\lambda_X \lambda_S}(g^2+y^2) \sin^2 2 \alpha + 6 g^4  \sin^4 \alpha- 6 y^4 \cos^4 \alpha \,,
\ee
from which eq.~(\ref{eq:CWpot}) also follows.
Let us note that while the $\beta$ function of $g$ is positive the one $\lambda_{XS}$ is always negative around the flat direction. 
This implies that the breaking of conformal invariance is driven by a marginally irrelevant (relevant) coupling when quartics (gauge) couplings dominate. 
As a consequence the RG improvement is more important when quartics dominate $\beta_{\lambda_{\rm eff}}$ as shown in the plots.

\begin{figure}[t]
\centering
\includegraphics[width=.46\textwidth]{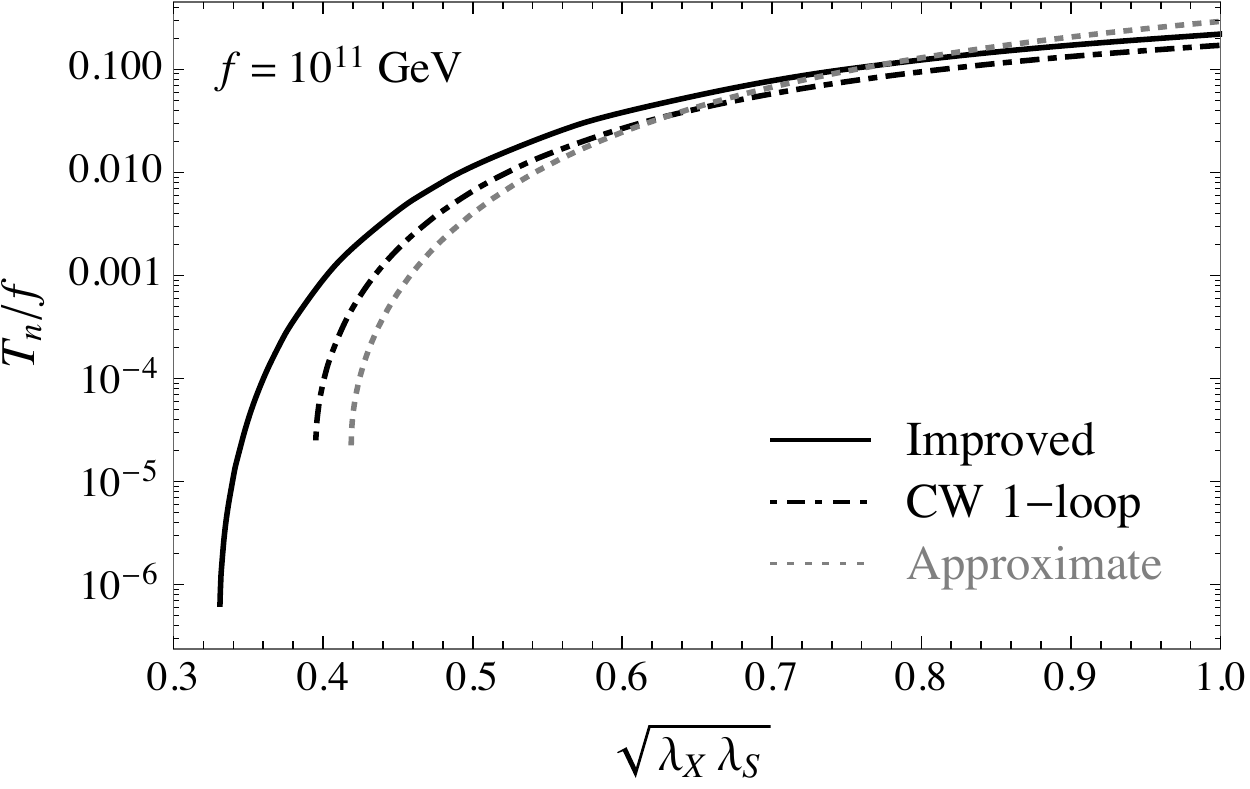}
\hspace{2em}
\includegraphics[width=.45\textwidth]{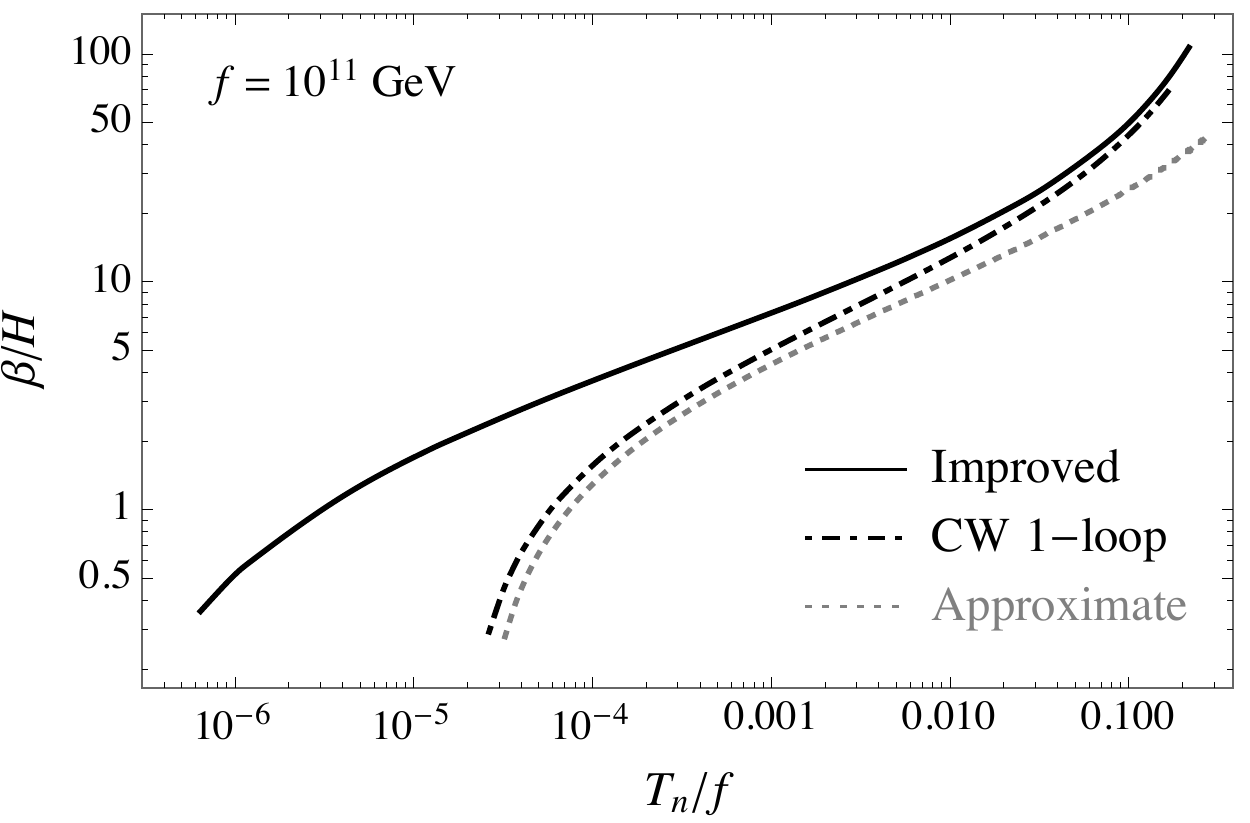} 
\caption{\it{Properties of the phase transition in the scenario with $g=0$, $y=0$ and $\lambda_S=\lambda_X$.
Left: Nucleation temperature versus coupling. Right: $\beta/H$ as a function of the nucleation temperature.
The solid curves correspond to the results using the improved 1-loop potential. The dot-dashed lines are instead obtained
from the usual 1-loop CW potential without improvement. Finally the dotted gray curves are derived through the analytical
approximation in eq.~(\ref{eq:an_approx_1}). The black lines include the full numerical thermal potential.}}
\label{fig:nucleation}
\end{figure}

\begin{figure}[t]
\centering
\includegraphics[width=.462\textwidth]{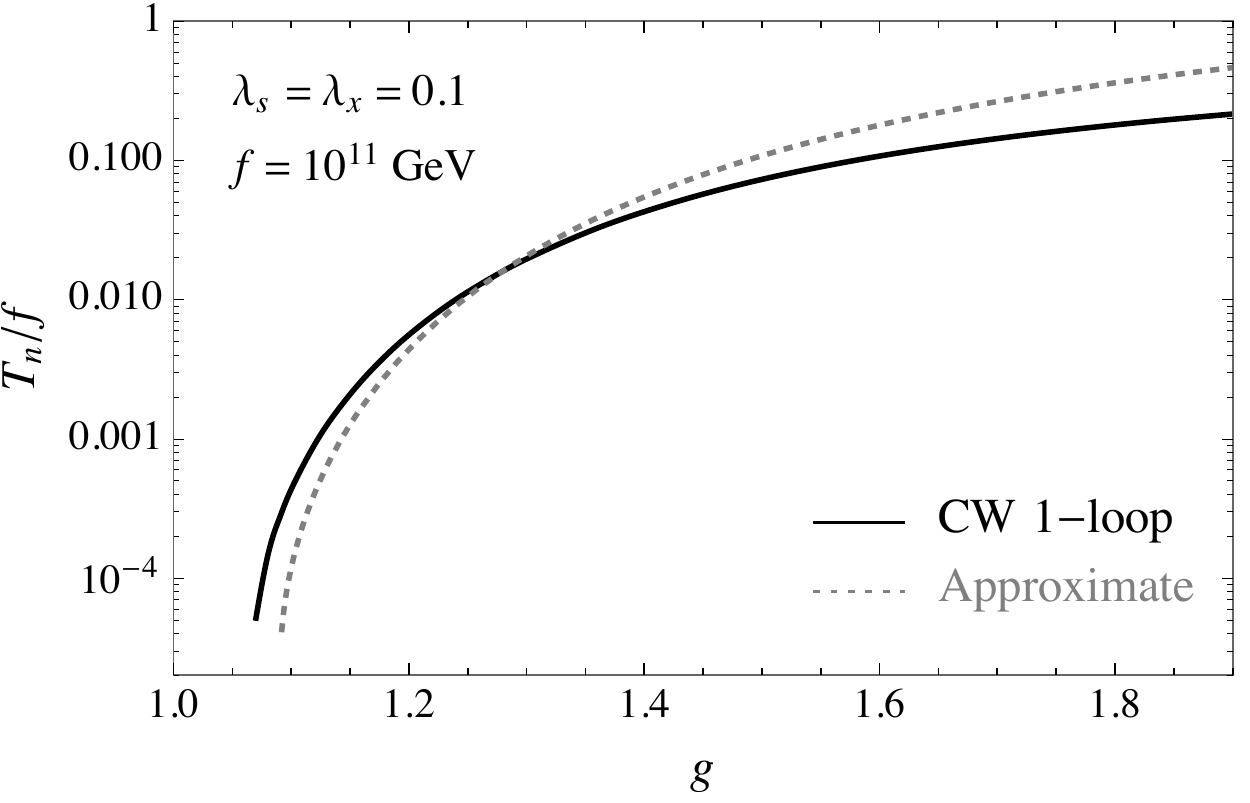} 
\hspace{2em}
\includegraphics[width=.45\textwidth]{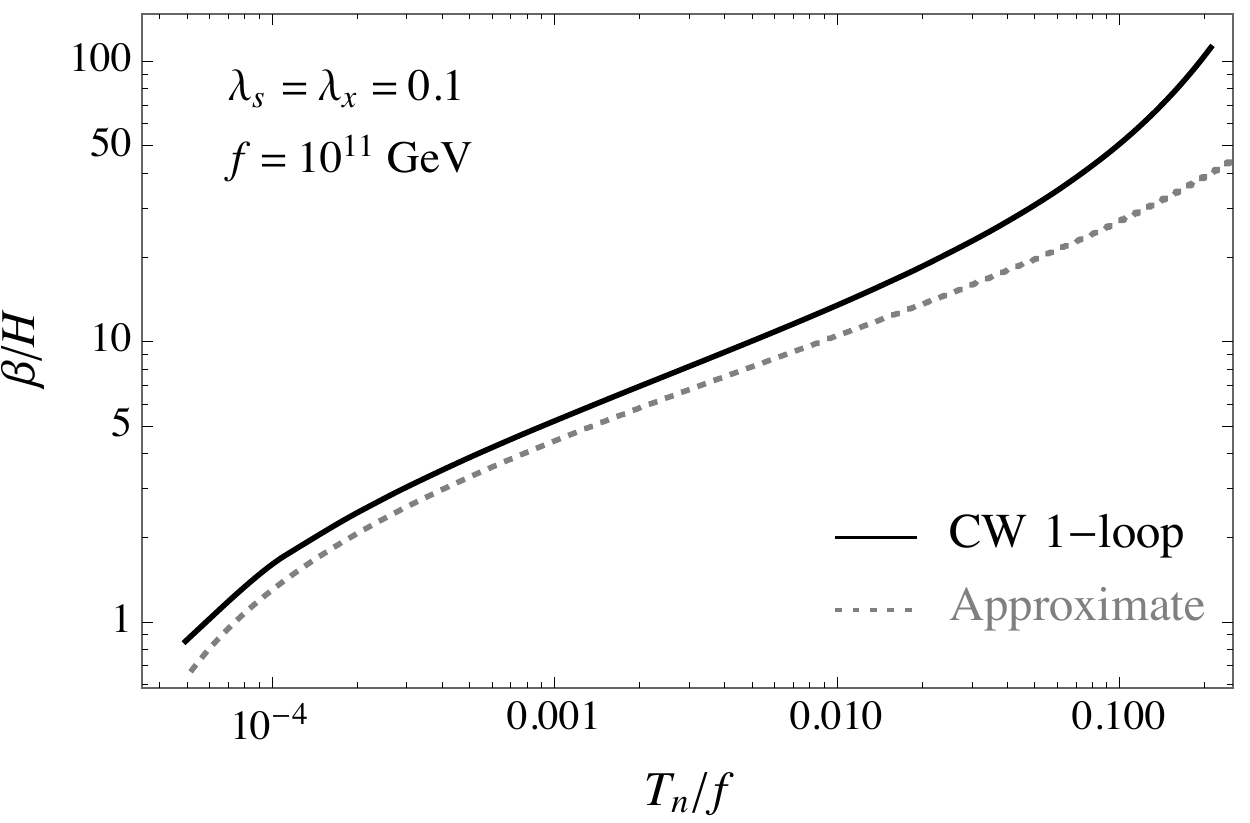} 
\caption{\it{Properties of the phase transition in the gauged scenario with $\lambda_S = \lambda_X\ll g^2$ and $y=0$. Left: Nucleation temperature versus coupling. Right: $\beta/H$ as a function of the nucleation temperature.
The solid curves are obtained using the 1-loop CW potential, while the dotted gray ones are derived through the approximate
analytic result. The black lines include the full numerical thermal potential.}}
\label{fig:nucleation_gauged}
\end{figure}

The computation of the parameters of the phase transition in the case of negligible gauging is shown in fig.~\ref{fig:nucleation}, for three different levels of approximation. It is visible how the RG-improvement of the effective potential gives different results in the region of extreme supercooling both for $T_n$ and the parameter $\beta/H(T_n)$. On the contrary in fig.~\ref{fig:nucleation_gauged}, where we assume dominance of the gauge contribution, we do not show the RG-improved potential, that we checked to be negligible.

 \paragraph{Reheating}

After completion of the phase transition the Universe is reheated. In the minimal scenario the only bridge between $X, S$ and the SM is provided 
by the coupling to colored fermions necessary to realize the QCD axion. The decay rate to the SM reads
\begin{equation}
\Gamma_\sigma= \frac 3 {8\pi}  y^2 \cos^2 \alpha M_\sigma\,,\qquad\quad\Gamma_\tau= \frac 3 {8\pi}  y^2 \sin^2 \alpha M_\tau\,.
\end{equation}
Assuming that the energy is carried by the light field $\sigma$, using eq. (\ref{eq:HI})  the reheating temperature is controlled by
\begin{equation}
\frac {\Gamma_\sigma}{H_I}=  \frac{3 \sqrt{3}}{2\pi} y^2 \cos^2 \alpha \frac {M_p}{f}\,,
\end{equation}
independently of  $\beta_{\lambda_{\rm eff}}$.

 \begin{figure}[t]
\centering
\includegraphics[width=.65\textwidth]{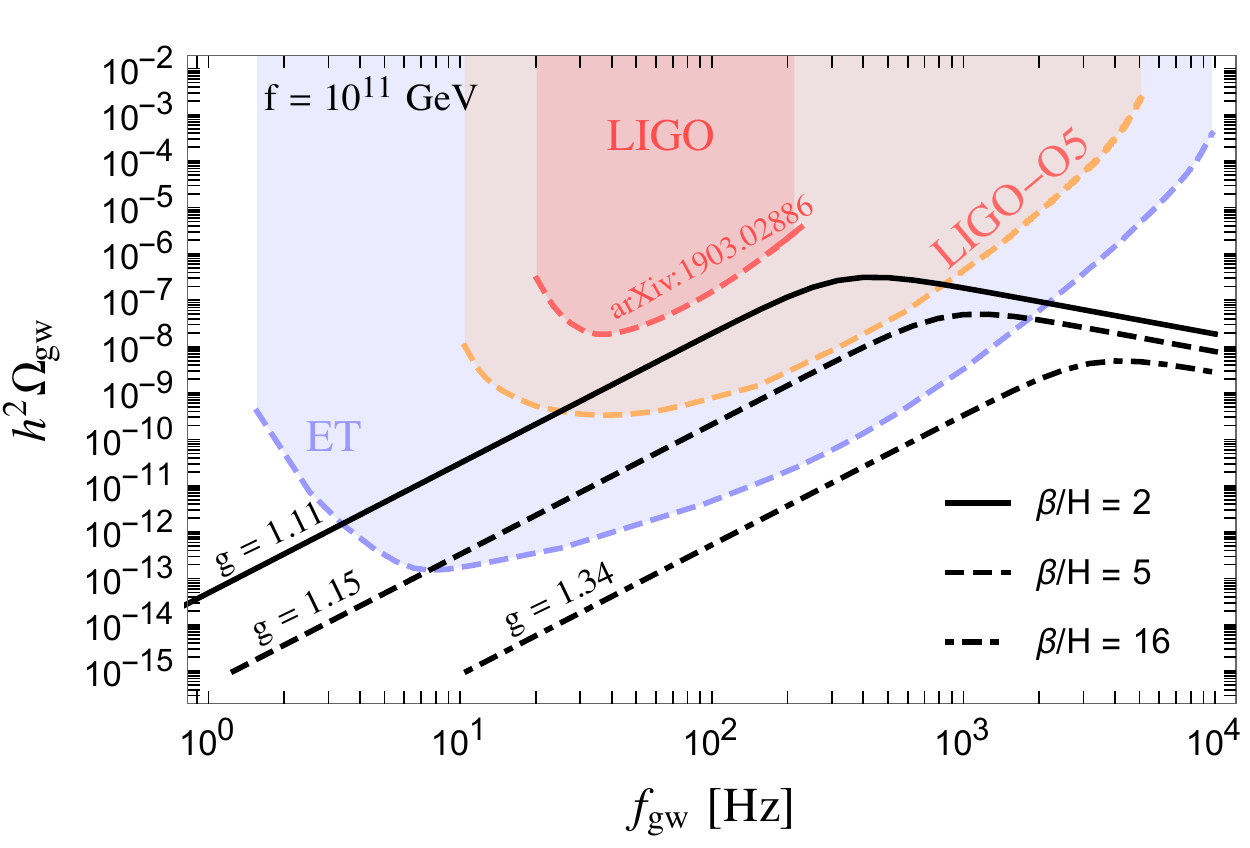}
\caption{\it{ Weakly coupled model. Predictions for the GW spectrum for three benchmark models in the gauge-dominance scenario. We also show the sensitivity curves of the LIGO (current bound \cite{LIGOScientific:2019vic} and projection of run O5) and the Einstein Telescope (ET) experiments. The same spectra can be realized in the purely quartic scenario with $g = 0$ and $\lambda_X =\lambda_S = (0.34, 0.38, 0.50)$, respectively.}}
\label{fig:gw-weakly}
\end{figure}

\paragraph{Gravitational-wave signals}
  In the presence of a large amount of supercooling the energy released while tunneling to the true vacuum can be much larger than the radiation energy and can dominate the energy of the Universe. 
 In such an empty Universe we only expect bubble collisions to be a source of GWs, while the effects of turbulence and sound wave propagation are subdominant. 
 Moreover, the effect of bubble collision is maximized during supercooling and reads~\cite{Caprini:2015zlo}, as a function of the frequency $f_{\rm gw}$,
  \bea
h^2  \Omega_{\rm gw}(f_{\rm gw}) \simeq  1.27 \times 10^{-6} \left( \frac{H(T_\textrm{RH})}{\beta}  \right)^2    \frac{3.8 (f_{\rm gw}/f_\textrm{peak})^{2.8}}{1 + 2.8 (f_{\rm gw}/f_\textrm{peak})^{3.8}}\,,
 \eea
 with the red-shifted peak frequency
 \bea
f_\textrm{peak} \simeq 3.83 \times 10^{2} \, \textrm{Hz}  \left( \frac{\beta}{H(T_\textrm{RH})}  \right)  \left( \frac{T_\textrm{RH}}{10^{10} \textrm{GeV}} \right) \,.
\eea

The GW spectrum depends on the temperature after the transition and reheating phases have completed. This generally does not coincide with the nucleation temperature, since only a small fraction of the energy 
released by the bubble goes into GWs (their production being Planck-mass suppressed).
Assuming a sufficiently fast reheating, $H(T_\textrm{RH}) \simeq H_I$, the relevant temperature for the GW spectrum can be simply estimated from the energy conservation condition
\bea
(1 - \Omega_{\rm gw}) (\Delta V + \rho_R(T_n)) = \rho_R(T_\textrm{RH})\,,
\eea
 which, for strong supercooling, simplifies to $T_\textrm{RH}^4 = 30/(\pi^2 g_*) \Delta V$. A large reheating temperature can shift the peak of the GW spectrum above the frequency regime where LIGO \cite{TheLIGOScientific:2014jea,Thrane:2013oya,TheLIGOScientific:2016wyq} and the Einstein Telescope (ET) \cite{Punturo:2010zz,Hild:2010id} have their optimal sensitivity.
On the other hand, the amplitude is completely controlled by $\beta/H(T_\textrm{RH}) \simeq \beta/H(T_n)$, which has been numerically computed and given in fig.~\ref{fig:nucleation}.
 
In fig.~\ref{fig:gw-weakly} we show the reach on stochastic gravitational background of the ET and LIGO observatories.
At present, the only existing bounds come from the LIGO collaboration~\cite{LIGOScientific:2019vic} from the combination of run O1 and O2. While this is not sufficiently strong to probe the models discussed here (assuming reasonable values of $\beta/H$), it is promisingly close to test these scenarios in the very near future. Indeed, at the end of the phase O5~\cite{TheLIGOScientific:2014jea}, LIGO would be already able to access part of the parameter space.

In order to get preliminary estimates we use the analysis developed in Ref.~\cite{Thrane:2013oya,Breitbach:2018ddu} and adopted in Ref.~\cite{Azatov:2019png} (see also \cite{Alanne:2019bsm}). From the knowledge of the effective noise strain $S_{\rm noise}(f_{\rm gw})$, as provided by the experimental collaborations, and assuming a power-law family of signals, one obtains the power-law integrated limit by maximising the  signal-to-noise ratio over the spectral index.\footnote{
The signal-to-noise ratio for a signal $\Omega(f_{\rm gw})$ is defined as
\be
\mathrm{SNR}=\sqrt{{\rm Time} \int_{f_{\rm min}}^{f_{\rm max}} df \bigg[\frac{\Omega(f)}{\Omega_{\rm noise}(f)}\bigg]^2}\,,   \qquad   \Omega_{\rm noise}(f_{\rm gw})=\frac{2\pi^2}{3H_0^2} f^3 S_{\rm noise}(f_{\rm gw})\,,
\ee
where the time is the integrated observational time, multiplied by the number of interferometers involved in the experiment.
A common practice for determining conservative bounds is to assume a power-law family of signals $\Omega_b(f_{\rm gw})=A_b f_{\rm gw}^b$. To extract the sensitivity then one can  find, at each frequency $f_{\rm gw}$, the largest value of $\Omega_b(f_{\rm gw})$ compatible with a given reference value of the signal-to-noise ratio, $\mathrm{SNR}_{\rm ref}$; i.e. we maximize over $b$ with the constraint that the test spectral density $\Omega_b(f)$ has a given $\mathrm{SNR}_{\rm ref}$ (here we take a value of 10). This gives a Power-Law Integrated (PLI) limit
\be
\Omega_{\rm PLI}(f_{\rm gw})= \max_b\,\, \Omega_b(f_{\rm gw})\big|_{\rm SNR_{ref}}= \max_b\,\, A_b\big|_{\rm SNR_{ref}} f_{\rm gw}^b = \frac{\mathrm{SNR}_{\rm ref}}{\sqrt{ \rm Time}} \max_b  \bigg[\bigg(\int_{f_{\rm min}}^{f_{\rm max}} d\bar f \frac{\bar f^{2b}}{\Omega_{\rm noise}^2(\bar f)}\bigg)^{-\frac{1}{2}}\, f_{\rm gw}^b\bigg] \,.
\ee
 }
 
Taking also into account the projected sensitivity of ET \cite{Sathyaprakash:2012jk}, one could be able to probe regions of the model with $g \lesssim 1.3$ ($\sqrt{\lambda_S \lambda_X} \lesssim 0.5$) characterised by nucleation temperatures $T_n/f \lesssim 10^{-2}$.

\section{Composite Axions}

We now turn to  scenarios where the axion is not an elementary field.
In this case the axion is a Nambu--Goldstone boson arising from the spontaneous breaking of the global symmetries of a strongly coupled dynamics 
that undergoes a confinement/deconfinement phase transition.
We consider two possible classes of models:
\begin{itemize}
\item Axion from SU($N$) gauge theories with massless elementary fermions charged under QCD \cite{Kim:1984pt}. In this context the QCD axion is the analog of $\pi_0$ in QCD and corresponds to a combination of phases  of the fermion condensates. 
\item Axion from a strongly coupled conformal (spontaneously broken) sector. At large-$N$ such a scenario is related to gauge theories in five dimensional AdS space through the 
AdS/CFT correspondence. In this realization the axion corresponds to the Wilson line of a 5D U(1) gauge field, and the anomalous coupling to gluons is realized through a Chern--Simons interaction
with SU(3) gauge fields. As we will see the PQ transition is intimately connected with the  breaking of conformal invariance. 
\end{itemize}

\subsection{Gauge theory axions}

In this class of models the axion appears as a Nambu--Goldstone boson of a confining gauge theory \cite{Kim:1984pt}. 
Such theories realize at low energy the KSVZ axions, and the PQ symmetry can be made accidental by 
appropriately engineering the gauge interactions to be chiral~\cite{1602.05427}.
Compared to weakly coupled models the axion has no radial mode, which, in practice, is replaced by the strong dynamics. 

In the simplest realization one considers an SU$(N)$ gauge theory with $N_F$ massless Dirac fermions. At least one of the fermions 
should be charged under QCD in order generate an anomaly, so that the minimal scenario requires $N_F=4$ (a color triplet and a singlet). 
Upon chiral symmetry breaking massless Nambu--Goldstone bosons are generated in the adjoint of the unbroken SU$(N_F)$ global symmetry. One can see that the symmetry of the SM singlet Nambu--Goldstone boson is anomalous under color and therefore realizes the QCD axion. 
The gauge dynamics is expected to have a first order phase transition for $3\le N_F< 4 N$ and $N\ge3$~\cite{Pisarski:1983ms}, and special cases have been verified on the lattice~\cite{Panero:2009tv}. 
We estimate the critical temperature as $T_c\sim f_\pi$ where $f_\pi$ is the decay constant of $SU(4)$ $\sigma-$model.
This is related to the axion decay constant by,
\begin{equation}
T_c\sim 2 A f_a
\end{equation}
where $A\delta^{ab}=2 N {\rm Tr}[T_{\rm PQ}T^a T^b]$ is the color anomaly where the flavor generators have 1/2 trace.

In this type of theories the phase transition however is not expected to lead to large supercooling. Indeed as soon as the temperature falls below the critical temperature 
the theory confines, exiting immediately from the scale invariant behavior.\footnote{A possible exception are confining gauge theories close to the conformal window 
estimated around $N_F\approx 4 N$ in QCD-like theories.} While it is presently not possible to compute the dynamics of the phase 
transition from first principles, estimates can be derived using effective models with the same symmetries of QCD~\cite{Bai:2018dxf,1904.07891,Croon:2019iuh}.
In Ref.~\cite{1904.07891} the parameters of the phase transition were estimated using Nambu--Jona-Lasinio and linear $\sigma$-models. While the details differ, 
the results indicate a nucleation temperature very close to the critical temperature and $\beta/H\ge 1000$. By applying these results to composite axions, 
for such parameters the amplitude of the GW signal  is suppressed  by plasma effects and the peak frequency is too large to be accessible at present experiments.

Let us also mention that these types of axion models, at least in their simplest realization, have domain wall number $N_{\rm DW}>1$ so that they contain stable domain walls. 
As a consequence it is most natural to consider these models when PQ symmetry is broken during inflation, such that the gravity wave spectrum is erased by the inflationary epoch.

\subsection{Conformal Models}\label{sec:strong}

A possible first order phase transition for the PQ sector can be triggered by the confining phase transition of large$-N$ conformal theories.
There is a vast literature on the conformal symmetry breaking in the context of the electro-weak scale starting with Ref. \cite{Creminelli:2001th}, see Refs.~\cite{Randall:2006py,Nardini:2007me,Konstandin:2011dr,vonHarling:2017yew,Bruggisser:2018mrt,Baratella:2018pxi,Agashe:2019lhy,Fujikura:2019oyi} for related work. 
Here we will consider a strongly coupled conformal sector with negligible couplings to the SM that triggers the PQ symmetry breaking, see  Refs. \cite{GarciaGarcia:2016xgv,Buchmuller:2019gfy} for other high scale realizations.
Not surprisingly the construction is similar in spirit to the one of massless elementary theories. 

In this context the phase transition is between a CFT at finite temperature and a spontaneously broken CFT with a light dilaton $\varphi$, 
the Nambu--Goldstone boson of the spontaneous breaking of scale invariance.
The model is defined at the UV scale $\Lambda$ by
\be\label{CFT}
\mathrm{CFT} + \frac{g}{\Lambda^{\epsilon}}\, \mathcal{O}\,,~~~~~~~~~~~~~|\epsilon| \ll 1\,.
\ee
The CFT is explicitly broken by a marginally relevant or irrelevant deformation $\mathcal{O}$ with dimension $4+\epsilon$ and also spontaneously broken.
Following the discussion in Ref.~\cite{Agashe:2019lhy} the explicit breaking of conformal invariance induces a slow evolution all the way down to the IR scale
that is captured by the running of the dilaton quartic coupling,
\be\label{lagrangian-normalized}
\mathscr{L}=\largeN \big[ (\partial \varphi)^2- \lambda(g(\varphi)) \varphi^4\big]= \largeN (\partial \varphi)^2 - \largeN \hat V(\varphi)\,.
\ee
The normalization of the kinetic term agrees with the dilaton being a glueball as in extra-dimensional realizations.

The explicit function $\lambda(g(\varphi))$ depends on how the CFT is explicitly broken, which in the generic parametrization of eq.~\eqref{CFT} 
is related to the running of the coupling $g$. In general the $\beta$-function in the large$-N$ limit has the structure
\be
\beta_g = \epsilon g + a\, N \frac{g^3}{16\pi^2} + \dots\,, \quad a\sim O(1)\,.
\ee
In the regime where $\epsilon > N g^2/(16\pi^2)$, the evolution of $g$ is dominated by the classical scaling dimension, while if $\epsilon\sim N g^2/(16\pi^2)$, the departure from scale invariance is the same as in the Coleman--Weinberg mechanism. Focusing on the classical evolution $g(\varphi)=(\varphi/\Lambda)^\epsilon\, g$, so that
\be
\lambda = \lambda_0 +  \lambda'( 0) \,g  \, (\varphi/\Lambda)^\epsilon + \dots,
\ee
which determines the effective potential of the dilaton through eq. (\ref{lagrangian-normalized}). By trading the product $\lambda'(0)g$ for the minimum of the potential $f$, one can write down
\be\label{dilaton-eff}
\hat V(\varphi)= \lambda_0 \varphi^4 \bigg[ 1 -\frac{4}{4+\epsilon} \bigg(\frac{\varphi}{f}\bigg)^\epsilon\bigg]+{\cal O}(\lambda_0^2)\,.
\ee
The potential has a minimum for $\lambda_0 \epsilon <0$. For the $\epsilon>0$ the evolution is controlled by a marginally irrelevant 
operator, while for $\epsilon<0$ the deformation is relevant and grows in the infrared.  Therefore for $\epsilon>0$ the breaking of conformal invariance
decouples in the IR. This is analogous to the discussion at the end of section 2.2.1.

Note that since $\varphi^\epsilon= 1+\epsilon \log \varphi + \dots $, for small enough $\epsilon$ this potential has the same structure as in the weakly coupled scalar models 
considered in eq.~\eqref{CW}, with the identification of $-\epsilon \lambda_0=N^2/(64 \pi^2)\beta_{\lambda_{\rm eff}}>0$.

Let us discuss the relevant normalizations. Because of the non-canonical kinetic term, the physical decay constant of the dilaton should
be identified with $f_d=N f /(4\pi) $. The axion decay constant is then  expected to be similar to $f_d$ up to order-one factors.
Actually in the extra-dimensional realization the axion scales as a meson while the dilaton as glueball. Large N countings would then 
indicate $f_a \sim f_d/\sqrt{N}$. We will neglect such factors in what follows and assume $f_a=f_d=N  f /(4\pi) $.

\bigskip
In order to connect the QCD axion to this sector one needs to simply assume that the CFT has a global symmetry U(1)$_{\rm PQ}\times$ SU(3), where the SU(3) factor 
is weakly gauged under QCD. The PQ symmetry should be anomalous under QCD. In operator language this means
\begin{equation}
\partial_\mu j^\mu_{\rm PQ}=  \frac {K}{16\pi^2} G_{\mu\nu}^a \tilde{G}^{a\, \mu\nu}\,,
\end{equation}
where $K$ is an integer.
We assume that when the CFT breaks it also breaks spontaneously the U(1) symmetry so that,
\begin{equation}
\langle 0 |j^\mu_{\rm PQ}(p) |a\rangle \sim \frac N {4\pi} f \, p^\mu  \,.
\end{equation}
Upon the spontaneous breaking of the PQ symmetry the axion degree of freedom acquires an anomalous coupling to gluons from the anomaly equation. 
It thus realizes the QCD axion. Because SM fermions have no PQ charge the low energy dynamics is the same as KSVZ models.

\paragraph{Extra-dimensional realization}

The above construction is dual, through the AdS/CFT correspondence, to  five-dimensional theories of gravity with negative cosmological constant.
From this point of view the phase transition corresponds to the Hawking--Page-type transition between AdS-Schwarzschild geometry 
and AdS with an IR brane \cite{Witten:1998zw,Creminelli:2001th}. 

For what concerns the dilaton the construction is the standard Randall--Sundrum scenario \cite{Randall:1999ee} with Goldberger--Wise stabilization \cite{Goldberger:1999uk}.
At zero temperature one considers AdS space with radius $L$ and metric
\begin{equation}
ds^2= \frac {L^2}{z^2} (dx^\mu dx_\mu + dz^2)\,.
\end{equation} 
The presence of an IR brane spontaneously breaks  conformal symmetry and generates a mass gap \cite{Rattazzi:2000hs}. 
The dilaton can be identified with the radion mode, whose potential  is in general quartic. 
The extra dimension can be stabilized at $z\gg L$ with the aid of the Goldberger--Wise field.
This requires the addition of an approximately massless scalar field, $\Pi$, dual to an almost marginal operator of the CFT,
with dimension $\Delta_\Pi=2+\sqrt{4+M_{\Pi}^2 L^2}$. For generic brane actions the 5D field acquires VEV and a potential for the radion field $\varphi$ is generated 
\be
V(\varphi)_{\rm GW}=\varphi^4\big[(4 + 2\epsilon)(v_1-v_0(\varphi/\Lambda)^\epsilon)^2-\epsilon v_1^2+\delta \big]\,,
\ee
where $\epsilon=\Delta_\Pi-4$, $v_{0,1}$ are the (normalized) values of the field $\Pi$ on the two boundaries of the 5D space.

This expression differs from eq.~\eqref{dilaton-eff}, as it includes an $(\varphi/\Lambda)^{2\epsilon}$ term, allowing for more general 
solutions. For example, for some choices of parameters the potential has a maximum between the origin and the minimum. 
We leave for the future the study of high scale phase transitions in holographic models.

\bigskip

For what concerns the axion, the construction is analogous to the one of AdS/QCD~\cite{DaRold:2005mxj,Erlich:2005qh} for chiral symmetry breaking, see also~\cite{Baratella:2018pxi}.
Through the AdS/CFT correspondence  CFT global symmetries are mapped into bulk gauge symmetries.
Since the CFT should have a global U(1) symmetry that is anomalous under QCD, the 5D action contains\footnote{The same 5D construction was discussed in Ref.~\cite{Choi:2003wr}.}
\begin{equation}
\mathscr{L}_5=-\frac 1 {4g_{\rm PQ}^2} F_{MN} F^{MN}- \frac 1 {4g_{3}^2} G_{MN}^a G^{a\,MN} +\frac {K}{192 \pi^2} \epsilon^{MNOPQ} A_M G_{NO}^a G^a_{PQ}+\dots\,,
\end{equation}
where $F$ and $G^a$ are the field strengths of U(1) and SU(3) 5D gauge fields and we have crucially included the Chern--Simons coupling
necessary to reproduce the anomalous coupling of the axion to gluons.

The action above must be supplemented by appropriate boundary conditions that produce massless 4D gauge fields for SU(3)
and an axion. These correspond to Dirichlet boundary condition for U(1)$_{\rm PQ}$ and Neumann for SU(3), 
\begin{equation}
A_\mu|_{z=L}=A_\mu|_{z=z_{\rm IR}}=0\,,~~~~~~~~~~~G_{\mu5}|_{z=L}=G_{\mu5}|_{z=z_{\rm IR}}=0\,.
\end{equation}
The axion corresponds to the Wilson line of $A_M$ in the fifth dimension. Its decay constant and low energy QCD coupling are given by
\begin{equation}
\frac 1 {g_s^2}=\frac 1 {g_0^2} + \frac L {g_3^2}\log \frac{z_{\rm IR}}L\,,\quad \quad f_a^2= \frac 2 {g_{\rm PQ}^2} \frac L {z_{\rm IR}^2}\,,
\end{equation}
where we allowed for a UV contribution $g_0$ to the QCD coupling. The Chern--Simons coupling induces the coupling of the axion with the QCD 
topological density, thus realizing a QCD axion model. Since the location of the IR brane is determined by the dilaton stabilization
mechanism, this setup realizes the conformal axion described above.

In reality, the boundary condition should be derived from the action principle. This can be done introducing a 5D scalar field charged under U(1)$_{\rm PQ}$
that acquires a VEV in the IR. If the field has mass $M_{\Phi}$ its profile in the extra dimension is
\begin{equation}
\Phi(z)\propto \frac 1 {z_{\rm IR}^{2\nu}-L^{2\nu}}\left[z_{\rm IR}^{2\nu} \left(\frac z {z_{\rm IR}}\right)^{2+\nu}-L^{2\nu}\left(\frac z {z_{\rm IR}}\right)^{2-\nu}\right]\,,\quad \quad  \nu=\sqrt{4+ M_\Phi^2 L^2}\,,
\end{equation}
which vanishes for $z=L$ to respect the $U(1)_{\rm PQ}$ symmetry.
Increasing the value of the mass the wave-function becomes more peaked in the IR. Since  $\Delta=2+\nu$ is the dimension of the dual operator,
breaking through boundary conditions is formally equivalent to an operator of large dimension. Note that in order not to 
affect the radion stabilization it is necessary that $\Delta >4$. In QCD-like theories instead the scalar corresponds to the relevant operator $\bar{q}_R^\alpha  q_L^\beta$,
so that no large hierarchy is generated.

At finite temperature two gravity solutions exist. One is  thermal AdS with the IR brane and the other is a black hole geometry where the  brane is replaced by the 
horizon. At high temperature the black hole solution has a smaller free energy and is thus favoured.
The boundary condition on the U(1) gauge field in this case respect the global U(1) symmetry and, as a consequence, the PQ symmetry is unbroken at high temperature.

\subsubsection{Phase transition}\label{sec:PTstrong}

Strongly coupled phase transitions are notoriously difficult to study because the degrees of freedom change across the transition, so that one cannot find an obvious trajectory in field space to describe the tunneling.  Interestingly, the problem can be circumvented when a light dilaton exists~\cite{Creminelli:2001th}. In this case the dynamics of the phase transition can be  described within the dilaton effective theory up to reasonable assumptions on the deconfined phase. Compared to the studies in the literature the main difference here is that the scale of our sector is set by $f_a$ and no phenomenological constraints on the anomalous dimensions apply.

At temperatures much higher than $\langle\varphi\rangle\equiv f$ we expect the system to be described by a hot CFT.  
The free-energy in this case is simply given by
\be\label{free-energy-CFT}
- F_{\rm CFT} (T\gg f) = b\, N^2 T^4\,,\qquad \quad b \sim O(1)\,.
\ee
Correspondingly, the  Hubble parameter in the false vacuum is
\begin{equation}
3 M_p^2 H^2= V_0+ \frac{g_* \pi^2}{30}T^4 \,,~~~~~~~~V_0= -\frac {N^2}{16\pi^2} \frac {\epsilon \lambda_0}{4+\epsilon} f^4\,,~~~~~~~~~~g_*= 106.75+ \frac{90}{\pi^2}\, b\, N^2\,.
\end{equation}

\begin{figure}[t]
\centering
\includegraphics[width=.95\textwidth]{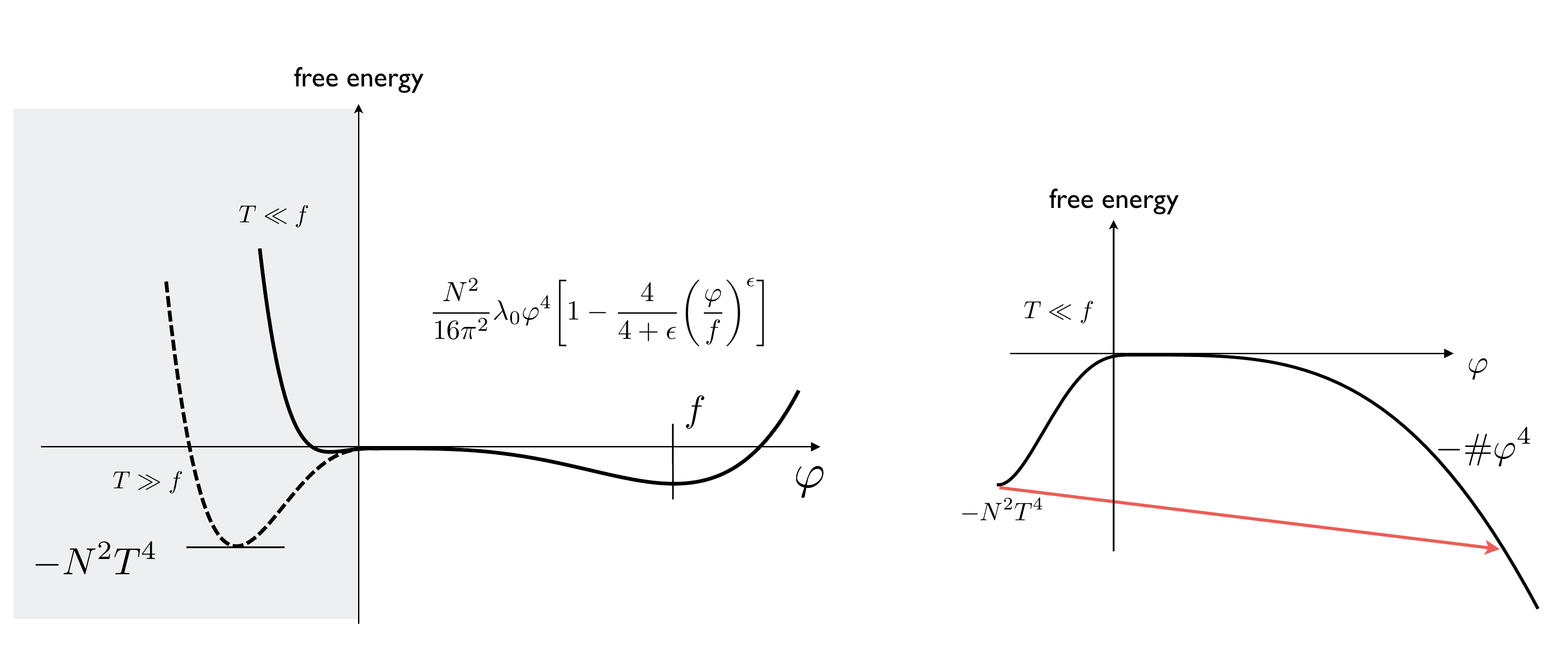} 
\caption{\it{On the left, free energy for different values of the temperature that affects the potential of the CFT phase. The universe is trapped in the false vacuum up to low temperatures where the relevant
region for tunneling is well approximated by a negative quartic potential (right panel).}}
\label{fig:cartoon_2}
\end{figure}

At temperatures below $f$ the confined phase has a lower free energy. This region is well described by the dilaton effective potential. 

In order to compute the tunneling rate various proposal have been proposed in the literature.
One possibility that we use in our numerical analysis is to calculate the bounce action by extending the dilaton potential to negative values, the sketch of the dynamics is shown in figure \ref{fig:cartoon_2}.
This amounts to continuing the potential for $\varphi < 0$ to connect smoothly 
to the value of the free-energy in eq.~\eqref{free-energy-CFT}. Various `guesses' have been considered in the literature for the shape of the potential. 
For example, as suggested by holography, the potential  can be extended to negative values of $\varphi$ as~\cite{Creminelli:2001th}
\be\label{black-hole}
\hat{V}_T(\varphi)=16\pi^2 b  \big( 4 \varphi^3 T + 3 \varphi^4 \big), 
\ee
whose minimum at $\varphi= - T$ is $-b N^2 T^4$ reproducing eq.~\eqref{lagrangian-normalized}.
Then one computes the bounce action with the standard boundary conditions: $\varphi'(0)=0$ and $\varphi(\infty)=-T$.
We have checked  that the result is weakly dependent on the choice of the potential. 

\medskip
An analytic approximation of the bounce action can be obtained by following Ref.~\cite{Konstandin:2010cd,Agashe:2019lhy}, see appendix A for details.
The bounce action can be split in an integral over the dilaton region and in another over the hot CFT. The first contribution can be estimated
computing the euclidean action of the dilaton with boundary conditions
\be\label{sundrum}
\varphi'(0)=0,\quad \varphi'(\varphi\to 0)=4\pi \sqrt{b} T^2.
\ee
The meaning of these boundary conditions is that, the dilaton should reach the origin with sufficient velocity to climb up the 
CFT free energy in the inverted potential. This  assumes negligible friction close to the origin, which is a good approximation at low temperatures. The full bounce action is then the sum of two contributions, $S_3^{(1)}$ from the pure dilaton EFT and $S_3^{(2)}$ from the thermal CFT.

The bounce $S_3^{(1)}$ can be estimated in a way similar to what we have done in section \ref{sec:weakcoupling}. 
In the region relevant for tunneling $\varphi\sim c T/f$ where $c= a |\lambda_0/(16\pi^2 b)|^{1/4}$ (best agreement with numerical results is obtained for $a=5$).
The dilaton potential can be approximated with a temperature-dependent quartic
\be\label{approximate-dilaton-potential}
\hat V(\varphi) \approx \lambda_0 \varphi^4 \bigg[1 - \frac{4}{4+\epsilon} \bigg(\frac{c T}{  f}\bigg)^\epsilon\bigg]\,.
\ee 
This approximation is more and more reliable in the limit of small $\epsilon$ and low temperatures, where the potential can be further simplified as
\be
\hat V(\varphi) \approx - |\epsilon  \lambda_0| \varphi^4 \log \bigg(\frac{f}{cT}\bigg)\,.
\ee
As shown in appendix A  with this approximation the bounce action is
\be
\frac{S_3^{(1)}}{T} = 28.5 \, \frac{N^2}{16\pi^2} \times \frac{  (16\pi^2 b )^{1/4}}{|\lambda_0\epsilon  \log (f/(cT)) |^{3/4}}\,.
\ee

\begin{figure}[t]
\centering
\includegraphics[width=.45\textwidth]{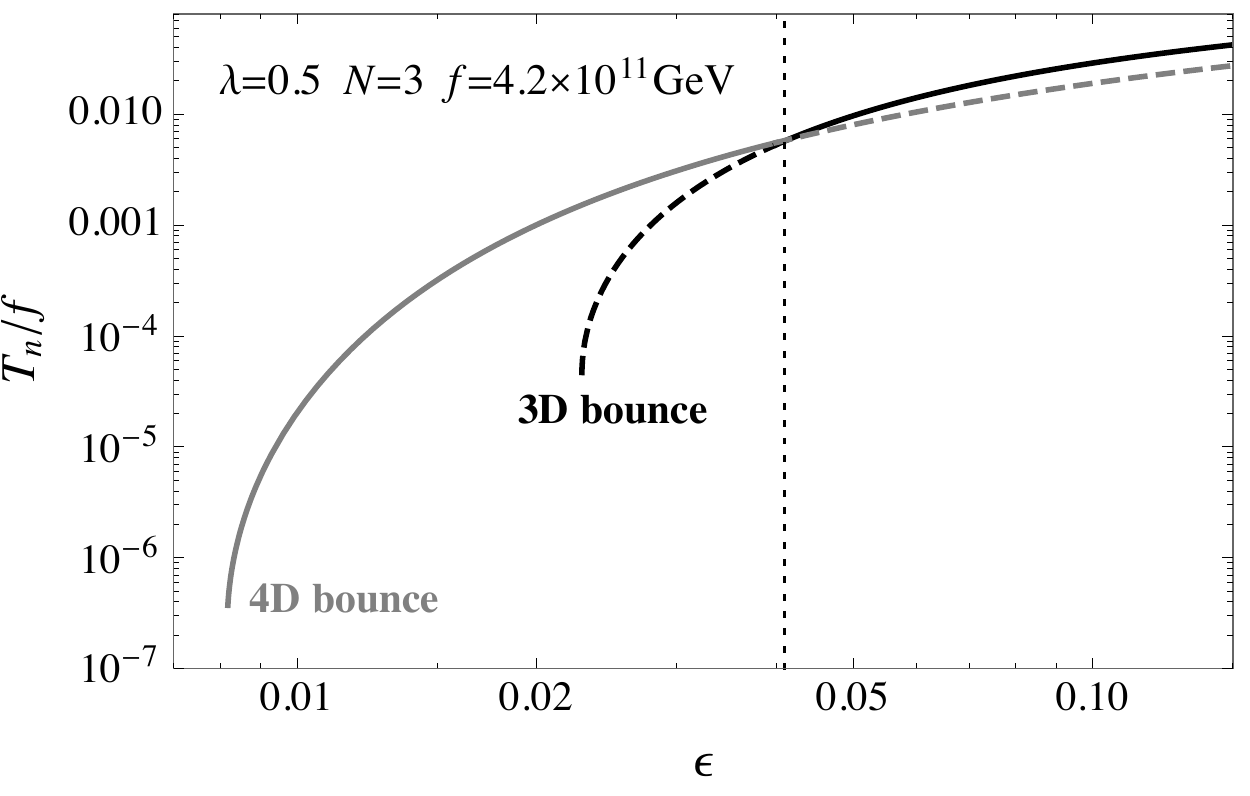} 
\includegraphics[width=.45\textwidth]{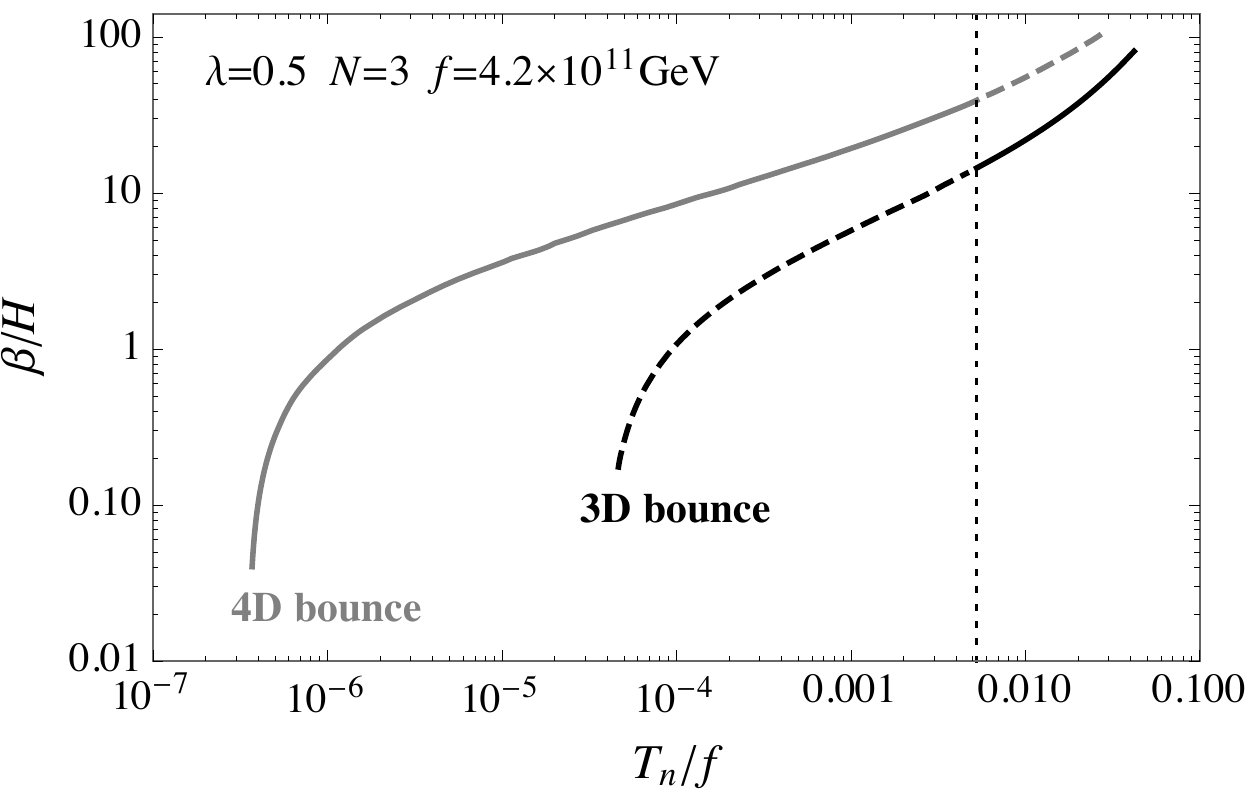} 
\caption{\it{Parameters of the phase transition in the strongly coupled case. Left: Nucleation temperature versus $\epsilon$. Right: $\beta/H$ as a function of the nucleation temperature. The normalization of the scale takes into account the factor $4\pi/N$.}}
\label{fig:PTCFT}
\end{figure}

The bounce $S_3^{(2)}$ can instead be computed in a thin wall approximation neglecting the friction as in \cite{Linde}
\begin{equation}
\frac {S_3^{(2)}}T= \frac {N^2}{8\pi^2 T} 4\pi R_*^2 \int_{-T}^0 \sqrt{\hat{V}_T(\varphi)}d\varphi \sim \frac {2\sqrt{b} N^2}{\sqrt{|\lambda_0|}}
\end{equation}
where $R_*^{-1}\sim |\lambda_0|^{1/4}T$ is  the critical bubble size.  Note that this contribution does not depend strongly on the details of the potential but only 
on the height and location of the minimum. This explains why our results are insensitive to the choice of potential. The O(3) bounce action can then be estimated as $S_3^{(1)}+S_3^{(2)}$. As emphasized in Ref.~\cite{Agashe:2019lhy}, due to the different scaling, 
for small $\lambda_0$ the tunneling is dominated by the dilaton contribution. In our scenarios we find that this contribution is not 
entirely negligible especially for what concerns the nucleation temperature.
This approximation  roughly agrees  with the exact numerical computation in figure \ref{fig:PTCFT}. 
The different parametric scaling from the weakly coupled case originates from the fact that the free energy  is dominated by $T^4$ in the deconfined regime, see appendix. 

We can repeat the derivation for the O(4) symmetric bounce. In that case
\be
S_4^{(1)} \sim 25\, \frac{N^2}{16\pi^2} \times \frac{1}{|\lambda_0\epsilon  \log (f/(cT))|}\,, \quad S_4^{(2)} \sim \frac{\pi \sqrt{b} N^2}{|\lambda_0|^{\frac 34}}\,.
\ee
The comparison of the 3d and 4d expressions already shows a difference with respect to the weakly coupled case of Section \ref{sec:weakcoupling}. Here, at low temperature we have dominance of O(4) bounces, that decrease faster with temperature by order $\sim \log(f/cT)^{1/4}$ as compared to the O(3) ones. This behavior is indeed confirmed in all of our numerical approaches.

\begin{figure}[t]
\centering
\includegraphics[width=.75\textwidth]{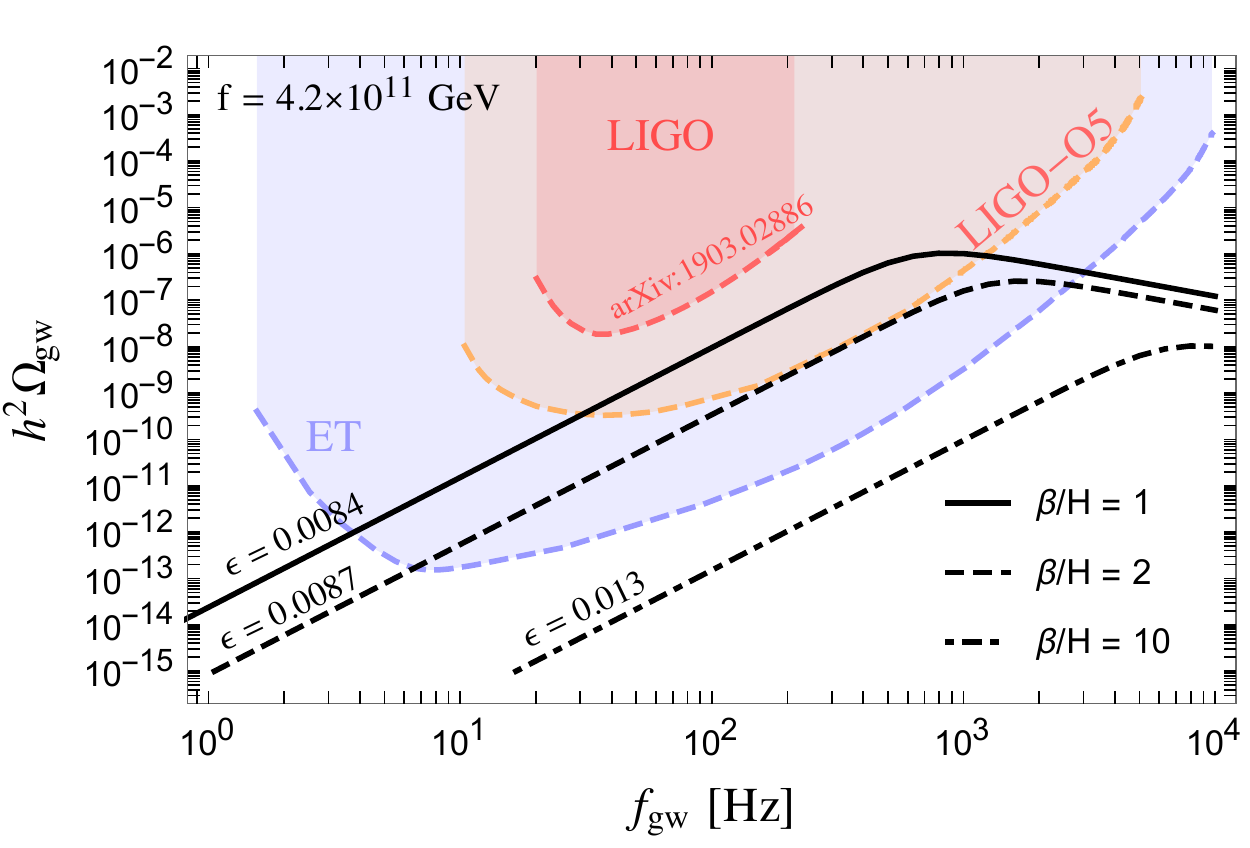}
\caption{\it{ Gravitational wave signal for strongly coupled conformal axions. The main difference compared to weakly coupled models (see figure \ref{fig:gw-weakly}) is the shift of the gravity 
wave peak to higher frequencies due to the larger dilaton decay constant.}}
\label{fig:gw-strongly}
\end{figure}

\paragraph{Nucleation temperature and gravitational wave signals}~\\
With the guidance of the analytic approximation it is then easier to understand the numerical results presented in fig.~\ref{fig:PTCFT}. In this figure we present the computation of the parameters of the phase transition $T_n$ and $\beta/H$. The nucleation temperature is determined again by using eq.~\eqref{Tn-all}, however, since in this model O(4) bounces are important, we computed the tunneling rate for both 3 and 4 dimensional bounces. In the case of the O(4) bounces, it is necessary to estimate the radius of the nucleated bubble that is then plugged into eq.~\eqref{gamma4}. We considered different definitions for the radius $R$: $i)$ $\varphi(R)=\varphi(r=0)/e$; $ii)$ $\varphi(R)=\varphi(r=0)/e^2$; $iii)$ $R= \int dr \varphi(r) r/( \int dr \varphi(r))$; and we found that they give results in excellent agreement with each other. The computation of the bounce action is largely insensitive to the choice of the potential, and we show the line  corresponding to the potential in eq.~\eqref{black-hole}.

From the numerical computations, we see that below $\epsilon\sim0.04$ the four dimensional tunneling starts to dominate. It is also in this region of parameter that $\beta/H$ can become appreciably smaller than O(10), thus enhancing the GW amplitude.  In fig.~\ref{fig:gw-strongly} we show the amplitude of the GW spectrum for three benchmark choices of $\beta/H=(1,2,10)$, which correspond to $\epsilon=(0.0084,0.0087,0.013)$. Even in this case the signal can be within the reach of the future upgrades of LIGO and future interferometers as the Einstein Telescope.

\section{Conclusions}

 \begin{figure}[t]
\centering
\includegraphics[width=.75\textwidth]{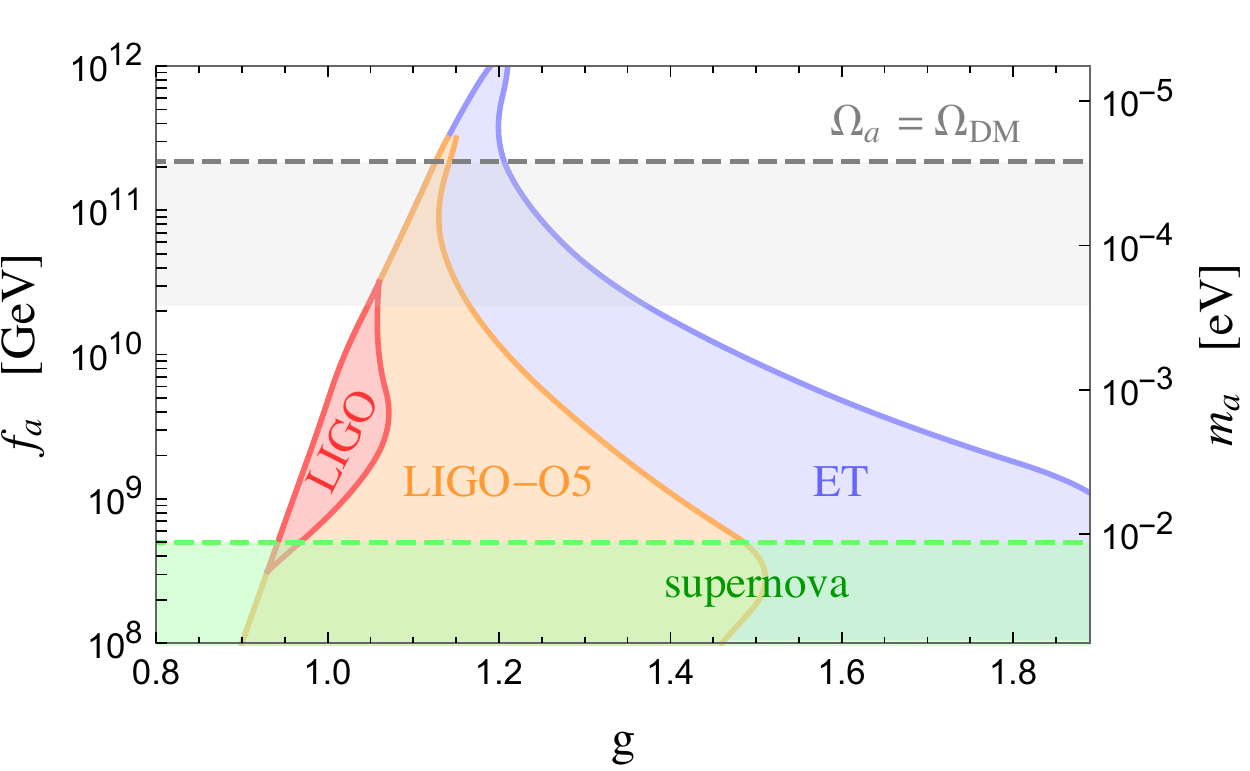}
\caption{\it{ Parameter space of the weakly coupled model for vanishing quartic couplings. Constraints on the QCD axion parameters arising from present and future GW interferometers and astrophysical bounds from supernovae are shown. The dashed gray line correspond to the pure misalignment contribution to axion DM, while the gray band represents the uncertainty due to the contribution from topological defects. }}
\label{fig:gw-weakly-exclusion}
\end{figure}

A recurrent dream is that the axion will be discovered in the near future. If the axion is the true solution to the strong CP problem, then the success of direct searches clearly depends on the advances of low-energy experiments. In this work we have shown that a new and complementary information on the physics of the QCD axion may also come from the study of GWs produced during the PQ phase transition. We find  that a detectable GW signal can be obtained even for the less favourable case where the axion is DM,  $f_a\approx 10^{11}$ GeV.

A stochastic GW signal requires the PQ phase transition to be first order.
While in the simplest KSVZ model the phase transition is second order,  we have found 
scenarios where the phase transition is first order\footnote{Different realizations of a first order PQ phase transition have been studied in Ref. \cite{Pomarol}. }. This is automatically realized if the theory is approximately conformal
either at weak or strong coupling. In this case the thermal phase transition can be very slow leading to a significant amount of 
supercooling. The supercooling maximizes the GW signal leading to an enhancement of the amplitude of the GW power spectrum
up to observable levels.

Our findings are generic, the phenomenological outcomes relevant for GWs depend little on the weakly or strongly coupled nature of the PQ model, since in both cases we find parameter-space regions that can be probed at LIGO or future ground-based interferometers. 
Of course there are some quantitative differences between the two scenarios, in particular on the type of tunneling that dominates the nucleation of bubbles at the phase transition. Indeed, we find in weakly coupled models of section~\ref{sec:weakcoupling} a preference for thermal tunneling (3D bounces), while in the strongly coupled models of section~\ref{sec:strong} a dominance of quantum tunneling (4D bounces) at low temperatures. We have however demonstrated that these differences play little role in the final GW signals. 

We wish to emphasize that even for the conservative choice $f_a=10^{11}$ GeV the GW signal might be detectable, as exemplified in figs.~\ref{fig:gw-weakly} and \ref{fig:gw-strongly}.
Today's GW frequency depends on the reheating mechanism right after the phase transition, and in the models explored in this paper the PQ-sector automatically reheats the 
SM through the coupling to gluons and colored fermions. If reheating is instantaneous, the peak frequency is in the range 100-1000 Hz which can be within the reach of the future stages of the LIGO experiment or future ground-based interferometers as the ET. 

One might wonder how the GW signals change as a function of $f_a$. Smaller values of $f_a$ might lead to axions that are not all the Dark Matter, or they can just be interpreted as axion-like particles. Clearly, allowing $f_a$ to take smaller values, the impact of GWs is bigger.  For $f_a< 10^{11}$ GeV a larger portion of the parameter space is within reach and some regions of the parameter space can even be excluded with present data from LIGO~\cite{LIGOScientific:2019vic}. This behavior is depicted in figure \ref{fig:gw-weakly-exclusion} where we consider the radiative PQ scenario with gauge dominance and we allow $f_a$ and the coupling $g$ to vary. It is interesting to notice the complementarity of GWs with existing bounds on the QCD axion parameter space.

\bigskip

There are many possible extensions of our work. For example one could consider more general deviations from conformal invariance.
In the weakly-coupled case this corresponds to adding masses for the elementary scalars, while in the strongly-coupled case to allow for more generic potentials as realized in holographic models.
Secondly the reheating process after supercooling is closely connected to the axion solution of the strong CP problem and could give informations of the spectrum of the theory. 
A slow reheating might lead to smaller reheating temperatures an thus smaller peak frequencies for the GW spectrum that are more easily detectable.
Finally our work can be generalized to study first order phase transition in other high scale models. We leave these and other questions to future work.

 \subsubsection*{Acknowledgments}
We thank Daniele Barducci, Francesco Bigazzi, Diego Redigolo, Filippo Sala and Daniele Teresi for useful discussions.
This work is supported by MIUR grants PRIN 2017FMJFMW and 2017L5W2PT, Ente Cassa di Risparmio di Firenze
and INFN grant STRONG.

\appendix

\section{Bounce in strongly coupled models}\label{app:bounce}

Different approaches for the computation of the bounce action for the strongly-coupled phase transitions appeared in the literature. 
Given the normalization adopted in the main text, the bounce equation is
\be
2\frac{d^2\varphi}{dr^2} +\frac{2(d-1)}{r}\frac{d\varphi}{dr}= \frac{\partial \hat V(\varphi)}{\partial \varphi}\,.
\ee

\paragraph{Method I}
This method merges two potentials 
\be
\mathscr{L}=\largeN (\partial \varphi)^2 - \largeN \bigg[\hat V_T(\varphi) \theta(-\varphi) + \hat V(\varphi) \theta(\varphi)\bigg]\,.
\ee 
The shape of $\hat V_T$ is largely unimportant as long as the potential satisfies a few requirements. It must be differentiable in the origin and minimized at $\varphi=-T$ with a value $\hat V_T\big|_{\rm min}=- 16\pi^2 b T^4$.\footnote{The exact position of the minimum is not of utmost importance, as far as it remains at a location $\varphi \sim -T$. We checked
numerically that modifying this relation by a factor of ${\cal O}(\textit{few})$ does not significantly affect the bounce action.}
Possible example for $\hat V_T(\varphi)$ are the following
\be
\hat{V}_T(\varphi)=-16\pi^2 b  \big( 2 \varphi^2 T^2 -\varphi^4 \big)\quad \mathrm{or} \quad\hat{V}_T(\varphi)=16\pi^2 b  \big( 4 \varphi^3 T + 3 \varphi^4 \big).
\ee
The bounce solution is then obtained in the usual way. It is convenient to rescale the field and distances as  $r\to z/(|\lambda_0|^{1/4} T)$ and $\varphi\to T \phi /|\lambda_0|^{1/4} $, so that we get
\be
\begin{split}
\frac{S_d}{T^{4-d}} &= \frac{2\pi^{d/2}}{\Gamma(d/2)}  \frac{ (16\pi^2b)^{\frac{4-d}{4}}}{|\lambda_0|^{d/4}}\largeN \times \\
& \int_0^\infty dz z^{d-1}\bigg[ \phi'^2 + \frac{\lambda_0}{|\lambda_0|} \phi^4 (1-\frac{4}{4+\epsilon} \left(\frac{T \phi}{|\lambda_0|^{1/4}\, f}\right)^\epsilon) \theta(\phi)+ \frac{V_T(T \phi /|\lambda_0|^{1/4})}{T^4} \theta(-\phi)+16\pi^2 b \bigg]
\end{split}
\ee
If $\hat{V}_T(\varphi)$ is a polynomial in $\varphi$ the only temperature dependent term in the above integrand is the part of the dilaton potential. The bounce solution corresponds to boundary values $\phi'(0)=0$ and $\phi(z=\infty)=-1$.

\paragraph{Method II}
In this case the boundary conditions when the field approaches the region close to the origin are as in eq.~\eqref{sundrum}.
After a rescaling $r\to z/(|\lambda_0|^{1/4} T)$ and $\varphi\to T \phi /|\lambda_0|^{1/4} $,  the boundary condition on the velocity close to the origin becomes
\be\label{bc-origin}
\frac{d\phi}{dz}\bigg|_{\phi\to 0}= 4\pi \sqrt{b} \,,
\ee
and the bounce integral can be written in general as
\be
\begin{split}
\frac{S_d}{T^{4-d}} = \frac{2\pi^{d/2}}{\Gamma(d/2)}  \frac{ (16\pi^2b)^{\frac{4-d}{4}}}{|\lambda_0|^{d/4}}\largeN  \int_0^{z_*} dz z^{d-1}\bigg[ \phi'^2 &+ \frac{\lambda_0}{|\lambda_0|} \phi^4 (1-\frac{4}{4+\epsilon} \left(\frac{T \phi}{|\lambda_0|^{1/4}\, f}\right)^\epsilon) +16\pi^2 b \bigg].
\end{split}
\ee
Notice that the extremum of integration $z_*$ is the time where the condition in eq.~\eqref{bc-origin} is satisfied. As discussed in section \ref{sec:PTstrong}, in order to gain some intuition on the size of the bounce action, one can write the potential in the region of tunneling as $ \lambda_0\,\varphi^4 \kappa(T/f,\epsilon)$, where $\kappa(T/f,\epsilon)$ is defined by matching with eq.~\eqref{approximate-dilaton-potential}.
With this approximation, by further rescaling $\phi\to (16\pi^2 b)^{1/4} \phi$ and $z\to z/(16\pi^2 b)^{1/4}$, we get
\be
\frac{S_d}{T^{4-d}} = \frac{2\pi^{d/2}}{\Gamma(d/2)}  \frac{(16\pi^2b)^{\frac{4-d}{4}}}{|\lambda_0|^{d/4}|\kappa(T/f,\epsilon)|^{d/4}}\largeN \int^{z_*} dz z^{d-1}\bigg[ \phi'^2 -   \phi^4  + 1 \bigg]\,.
\ee
We then simply need to evaluate the integral with appropriate boundary conditions for the solution
$$A_d\equiv\int^{z_*} dz z^{d-1}\bigg[ \phi'^2 -  \phi^4  + 1 \bigg]\quad \phi'(0)=0, \phi'^2|_{\phi=0}=1,$$
where $A_{3}=2.268$ and $A_4\approx 1.3$.
We get the following estimates for O(3) and O(4) bounces
\be
\frac{S_3}{T} = 28.5 \, \frac{N^2}{16\pi^2} \times \frac{  (16\pi^2)^{1/4} \, b^{1/4}}{|\lambda_0|^{3/4}}\times \frac{1}{| \kappa(T,\epsilon)|^{3/4}},\qquad S_4 = 2\pi^2 A_4\, \frac{N^2}{16\pi^2} \times \frac{1}{|\lambda_0|}\frac{1}{|  \kappa(T,\epsilon)|}\,.
\ee

\section{Plots of the extended parameter space}\label{app:plots}

\begin{figure}[t]
\centering
\includegraphics[width=.45\textwidth]{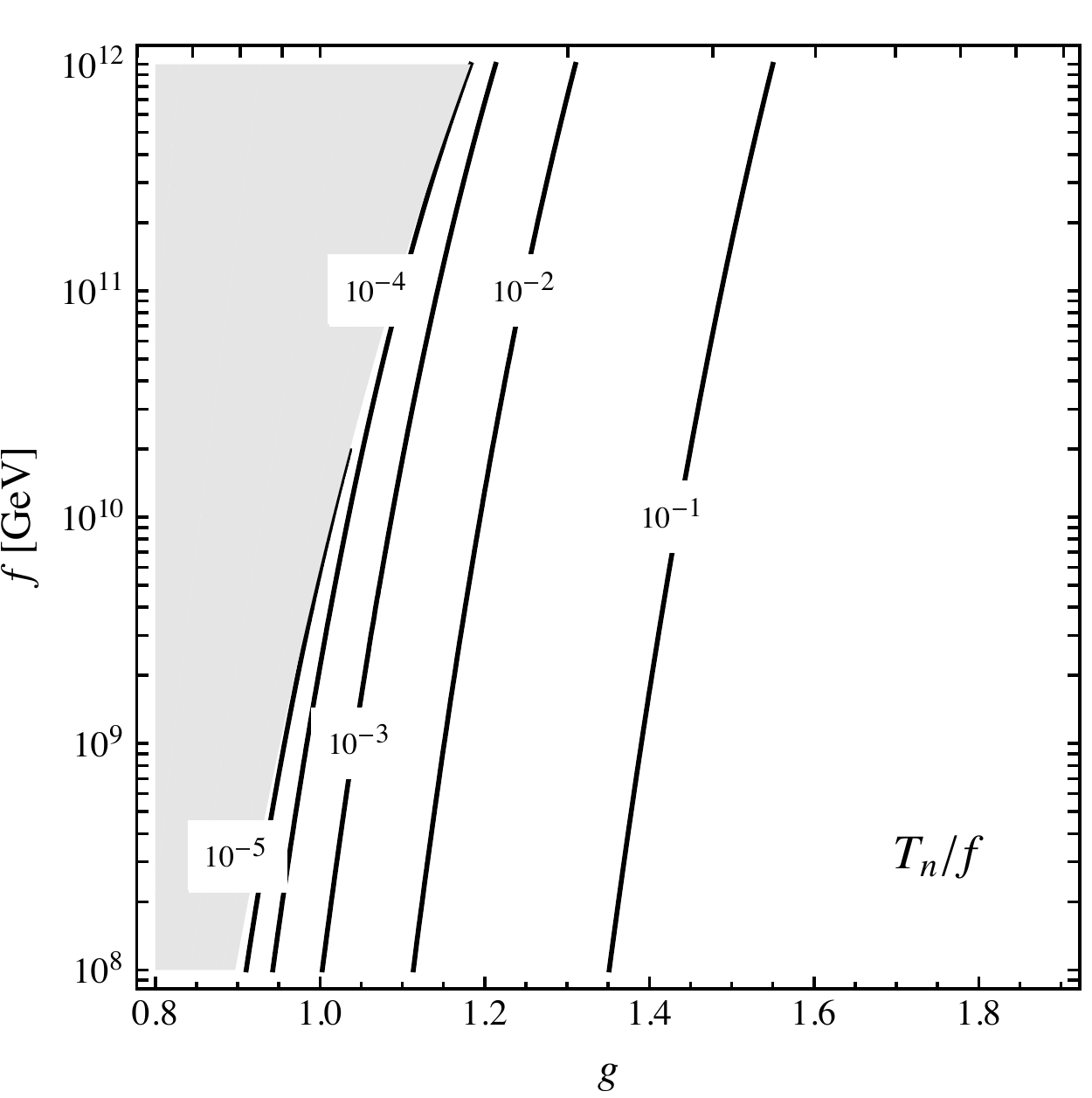} \quad \quad
\includegraphics[width=.45\textwidth]{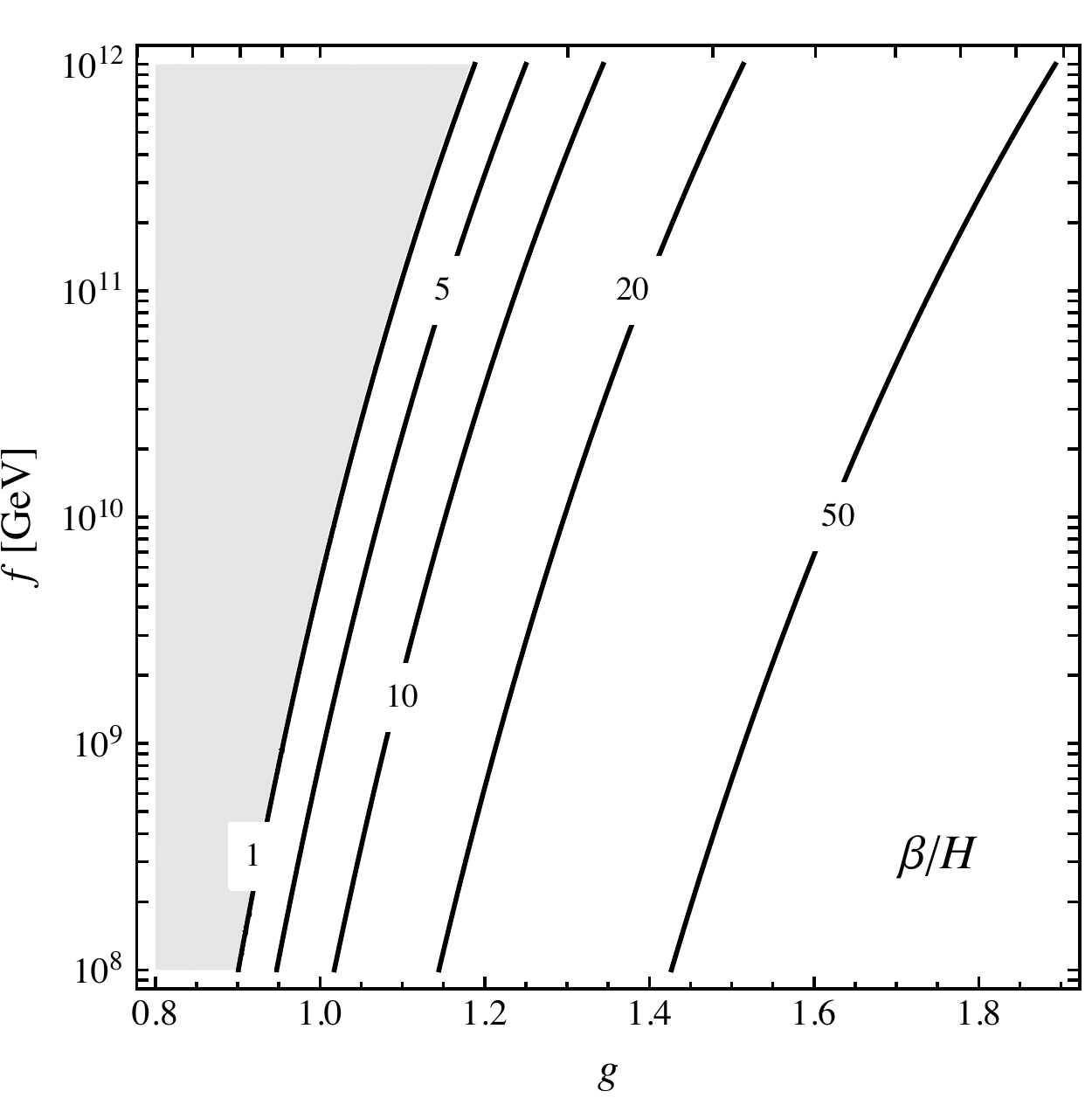}
\caption{\it{ Contour lines of the the nucleation temperature $T_n/f$ (left plot) and the $\beta/H$ parameter (right plot) in the gauged weakly coupled model. In the gray regions nucleation never happens.}}
\label{fig:contours}
\end{figure}

Here we provide more details on the parameters of the phase transitions in the weakly coupled scenarios where the gauge contribution is dominant. 
In fig. \ref{fig:contours} we show the behavior of both the normalized nucleation temperature and the $\beta/H$ parameter as functions of the gauge coupling $g$ and the scale $f$, 
focusing on the region of large supercooling.
The dependence in the relevant part of the parameter space follows the analytic approximation discussed in the text and shown in eq. (\ref{witten}) and (\ref{beta-H}). 
The approximation is particularly appropriate in the region with large supercooling, with $T_n/f \lesssim 10^{-2}$.
For a fixed value of the gauge coupling, a larger amount of supercooling (and, thus, a smaller $\beta/H$) can be achieved with a larger scale $f$ since the nucleation temperature is dominated by the exponential factor of eq. (\ref{witten}).

\end{document}